\documentclass[fleqn,usenatbib]{mnras}
\usepackage{newtxtext,newtxmath}
\usepackage{lineno}
\usepackage[T1]{fontenc}
\usepackage{adjustbox}
\usepackage{blindtext}
\usepackage{multirow} 
\DeclareRobustCommand{\VAN}[3]{#2}
\let\VANthebibliography\thebibliography
\def\thebibliography{\DeclareRobustCommand{\VAN}[3]{##3}\VANthebibliography}

\usepackage{booktabs} 
\usepackage{comment}
\usepackage{ulem}
\usepackage{graphicx}	% Including figure files
\usepackage{amsmath}	% Advanced maths commands
\usepackage{dcolumn}
\usepackage{hyperref}
\usepackage{kantlipsum}
\usepackage{dblfloatfix}
\usepackage{longtable}%https://www.overleaf.com/project/628b33c17a2ab65a09ea4734
\usepackage{array}
\usepackage{caption}
\usepackage{threeparttable}
\usepackage{url}
\usepackage{subcaption}
\newcolumntype{M}[1]{>{\centering\arraybackslash}m{#1}}
\newcolumntype{N}{@{}m{0pt}@{}}

\usepackage{bm}

\title [Disk Evolution in GX 339–4] {Accretion Disk Evolution in GX 339–4 Across Spectral States Using NuSTAR, NICER, and \textit{Insight}-HXMT Observations}

\author[Dhaka et al.]{Ruchika Dhaka,$^{1}$\thanks{E-mail: ruchika@iitk.ac.in, ruchikadhaka1997@gmail.com}
Ranjeev Misra,$^{2}$
and Suraj Kumar Chaurasia$^{3}$
\\
$^{1}$Department of Physics, IIT Kanpur, Kanpur, Uttar Pradesh
208016, India\\
$^{2}$Inter-University Center for Astronomy and Astrophysics,
Ganeshkhind, Pune 411007, India\\
$^{3}$Banaras Hindu University, Varanasi, Uttar Pradesh 221005, India 
}

\date{Accepted XXX. Received YYY; in original form ZZZ}

\pubyear{2025}

\begin{document}
\label{firstpage}
\pagerange{\pageref{firstpage}--\pageref{lastpage}}
\maketitle

\begin{abstract}
We present a broadband spectral analysis of the black hole X-ray binary GX 339--4 during its 2021 outburst, covering both hard and soft spectral states. Using simultaneous observations from \textit{NuSTAR}, \textit{NICER}, and \textit{Insight--HXMT}, we investigate the evolution of the accretion disk with a focus on the disk normalization derived from the \texttt{diskbb} component, which serves as a proxy for the apparent inner disk radius. In the standard single Comptonization model, the disk normalization in the hard state is more than an order of magnitude lower than in the soft state ($\sim$0.3~$\times$~10$^3$ vs. $\sim$3.0~$\times$~10$^3$). This result contradicts the widely accepted view that the disk radius is smaller in the soft state than in the hard state. By incorporating an additional warm Comptonization component, the disk normalization in the hard state increases to values ($\gtrsim 10^4$) exceeding those in the soft state ($\sim 10^3$), yielding results consistent with a physically truncated, cooler accretion disk. The results of this work support the presence of a dual-corona geometry in the hard state, comprising both a hot, optically thin corona and a warm, optically thick corona, while the soft state spectrum is well described by a single hot Comptonization component alone. Our findings emphasize the importance of including a warm corona in hard-state spectra, as it leads to a more physically consistent picture of the accretion geometry across spectral states.
\end{abstract}


%%%%%%%%%%%%%%%%% BODY OF PAPER %%%%%%%%%%%%%%%%%%

\section{Introduction}
\label{sec:intro}
Black hole X-ray binaries (BHXBs) undergo dramatic changes in their X-ray emission during outbursts, typically traced on a hardness-intensity diagram (HID) \citep[][]{Miyamoto1995, Fender2004, Belloni2005}. Outbursts generally begin in a low hard state (LHS), where the X-ray spectrum is dominated by a hard power-law component produced via inverse Compton scattering by a hot corona near the black hole \citep[][]{SunyaevTitarchuk1980, HaardtMaraschi1993, DoneGierlinskiKubota2007}. As the outburst progresses, the source transitions through intermediate states to a high soft state (HSS), where thermal emission from the accretion disk dominates. This transition typically occurs within days, with spectral softening as the inner edge of the disk moves inward. During the decline, the system returns to the LHS, tracing a hysteresis loop in the HID. These transitions reveal changes in accretion geometry, particularly in the inner disk radius \citep[][]{2004MNRAS.355.1105F, 2006AdSpR..38.2801B, 2010LNP...794...53B}.

The distance to GX~339$-$4 and the black-hole mass remain uncertain; optical and near-infrared observations provide robust lower limits of $D \gtrsim 5$–6~kpc \citep{Hynes2003,Heida2017}, while evolutionary modeling allows a broader distance range extending to $\sim$8–12~kpc \citep{Zdziarski2019}. The black-hole mass is typically estimated to be $M_{\rm BH} \sim 9$–12~$M_{\odot}$, with broader allowed ranges depending on the method and assumptions \citep{Parker2016,Sreehari2019,Zdziarski2019}. The accretion flow of the system is dominated by Roche lobe overflow from its low-mass companion star. The inclination angle of the accretion disk has an intermediate value $\approx 40^\circ$– $60^\circ$ \citep[][]{2019MNRAS.488.1026Z, 2015ApJ...808..122F}. Studies have found that the black hole in GX 339-4 has a very high spin ($\sim0.95$) value \citep{2015ApJ...813...84G, 2016ApJ...821L...6P}. GX~339$-$4 has exhibited both successful outbursts, which follow the canonical sequence of accretion states, and failed (hard-only or incomplete) outbursts, in which the source does not reach the soft state (e.g. \citealt{Bhowmick2021}; \citealt{Garcia2019}). Strong relativistic reflection features, including the broad iron $\text{K}\alpha$  line, have been observed in both hard and soft states, providing insights into the spin of the black hole and the geometry of the inner accretion disk \citep[][]{2004ApJ...606L.131M, 2015ApJ...813...84G, 2022MNRAS.513.4308L, 2023ApJ...950....5L}. 

The inner radius of the accretion disk plays an important role in understanding the accretion process, particularly during state transitions. The inner edge of the disk is believed to approach the innermost stable circular orbit (ISCO) as the system transitions from the hard to the soft state \citep{1997ApJ...489..865E,RemillardMcClintock2006,DoneGierlinskiKubota2007, 2023MNRAS.524.2721D}. This behavior, commonly observed in BHXBs, corresponds to an increasing accretion rate, where the disk appears truncated at larger radii during the LHS and moves inward during the HSS. 
However, the exact point at which the disc reaches the ISCO and the mechanism that triggers the state transition remains unclear. In some cases, measurements of the size of the inner edge have been controversial and highly dependent on the choice of models \citep[][]{2019MNRAS.485.3845D, 2019MNRAS.486.2137M, 2021ApJ...906...69Z}.
The relativistic reflection component, arising from coronal emission reflecting off the optically thick accretion disk near the black hole, offers a powerful probe of the spacetime properties and geometry of the accreting system \citep[][]{2014SSRv..183..277R, 2017RvMP...89b5001B, 2019PhRvD..99l3007L, 2021SSRv..217...65B}. A key characteristic of this component is the broadened iron line around 6.4 keV, sculpted by relativistic effects near the black hole such as Doppler boosting, light bending, and gravitational redshift \citep[][]{2000PASP..112.1145F, 1989MNRAS.238..729F, 2010MNRAS.409.1534D}. This broadened line feature is crucial for mapping the innermost regions surrounding the black hole \citep[][]{1989MNRAS.238..729F}.

\citet{2020ApJ...904..201S} studied two distinct outbursts of GX 339-4 at different luminosities, using a single comptonization model to explain the spectral evolution. Their findings indicated that, in the bright hard state, the inner edge of the accretion disk remained very close to the innermost stable circular orbit (ISCO) \citep[][]{2015ApJ...813...84G, 2019ApJ...885...48G, 2018ApJ...855...61W}, regardless of the luminosity at which the black hole transitioned from the bright hard state to the soft state. By fitting the spectra with the RELXILL family of relativistic reflection models, they concluded that the inner edge of the accretion disk reached around 9 $R_g$ by the onset of the bright hard state and remained at approximately 3 $R_g$ during the intermediate state transition. In a similar study, \citet{2023ApJ...950....5L} also examined the transition of GX 339-4 from the hard to soft state. Their results reaffirmed that the inner disk radius remains close to the ISCO throughout the transition of X-ray spectral states. \citet{2016MNRAS.458.2199B} studied GX 339-4 in the hard state using XMM-Newton data. Their analysis, utilizing the relativistic reflection model ({\fontfamily{pcr}\selectfont relxill}\footnote{\label{ft3}\href{https://www.sternwarte.uni-erlangen.de/~dauser/research/relxill}{https://www.sternwarte.uni-erlangen.de/~dauser/research/relxill}}), explored various models including Comptonization and lamp-post geometry. In their coronal models, the inner radius was consistently found to be high in the hard spectral state. Furthermore, when applying the lamp-post model, \citet{2016MNRAS.458.2199B} found the inner radius to become unconstrained. However, fixing it to the ISCO resulted in a large coronal height, implying weak relativistic broadening. 

\citet{2015ApJ...808....9P} analyzed Cyg X-1 in the hard state and determined the inner disk radius to be around $\sim 3 r_g$. However, a subsequent re-analysis by \citet{2017MNRAS.472.4220B} proposed a more complex coronal geometry, involving two distinct Comptonizing regions. This re-analysis revealed that the inner disk is truncated in the hard state, with a radius between 13–20 $r_g$. \citet{2013PASJ...65...80Y} studied Cyg X-1 data from Suzaku in hard state using temporal spectroscopy, concluding that both hard and soft Comptonization components are necessary for an accurate spectral fit. Black hole spin measurements using the disk continuum method for LMC X-1 and Cyg X-1, yielded different results when incorporating single versus double Comptonization models. The inclusion of an optically thick Comptonizing layer significantly reduced the spin parameter \citep[][]{2024ApJ...962..101Z}. Thus, it is important to estimate the inner radius of black hole systems during different spectral states to understand the geometry of the system. In a recent study, using data from NICER and Insight-HXMT, \citet{Li2025} have shown that for the 2018 outburst of  MAXI J1820+070 the inner disk radius reached a minimum of approximately 4.5 $r_g$ during the intermediate state (IMS), which is smaller than the radius measured in the soft state, where it remained around 9 $r_g$. Such inconsistencies from the standard picture needs to be verified using broadband spectral analysis of sources in different states. 

In this study, we analyze simultaneous observations of GX 339–4 from NuSTAR, NICER, and Insight–HXMT, covering both hard and soft spectral states. The combined data offer broad energy coverage, which helps constrain the spectral components more accurately. In particular, we focus on the evolution of the disk normalization, which serves as an indicator of the apparent inner disk radius and provides insight into changes in the accretion geometry across different states. The observation and data reduction procedures are described in Section \ref{sec:obs_data_red}. The results of our spectral fitting are presented in Section \ref{sec:result} , and discussion of our findings follows in Section \ref{sec:discussion}.

\begin{table*}
    \centering
        \setlength{\tabcolsep}{8pt} % Adjust column spacing
         \caption{NICER, NuSTAR and HXMT observations of GX 339-4 during the 2021 Outburst. For HXMT, the listed exposure time is for the LE instrument. For Nustar the mentioned exposure is for FPMA and for NICER the mentioned exposure is for XTI.}
        \label{tab:table1}
        \begin{tabular}{c c c c c c c c} % Adjusted column formatting
        \toprule
        & & \multicolumn{2}{c}{\textbf{NICER}} & \multicolumn{2}{c}{\textbf{NuSTAR}} & \multicolumn{2}{c}{\textbf{HXMT}} \\
        \cmidrule(lr){3-4} \cmidrule(lr){5-6} \cmidrule(lr){7-8}
        Epoch & Date & Obs ID & Exposure (s)  & Obs ID & Exposure (s) & Obs ID & Exposure (s) \\
        \midrule
        1 & 2021-02-20  & 3558010501 & 2429 & 90702303005 & 23420 & P030409300101 & 829 \\ [6pt]
         & ... & ... & ... & ... & ... & P030402400301 & 1418\\ [6pt]
        2 & 2021-03-08 & 3558010903 & 11820 & 90702303007 & 23420 & P030409301901 & 1830 \\ [6pt]
         & ... & ... & ... & ... & ... & P030402402201 & 180 \\ [6pt]
        3 & 2021-03-26 & 4133010103 & 4555 & 90702303009 & 15790 & P030402403801 & 2148 \\ [6pt]
         & ... & ... & ... & ... & ... & P030402403802 & 958 \\ [6pt]
         & ... & ... & ... & ... & ... & P030402403803 & 539 \\ [6pt]
        4 & 2021-04-01 & 4133010109 & 5059 & 90702303011 & 14750 & P030402404305 & 837 \\ [6pt]
         & ... & ...  & ... & ... & ... & P030402404306 & 1017 \\ [6pt]
         & ... & ... & ... & ... & ... & P030402404308 & 955 \\ [6pt]
         & ... & ... & ... & ... & ... & P030402404309 & 239 \\ [6pt]
        5 & 2021-04-24 & 4133010131 & 2041 & 90702303013 & 20140 & P030402406001 & 4416 \\ [6pt]
          & ... & ... & ... & ... & ... & P030402406002 &  2179 \\ [6pt]
           & ... & ... & ... & ... & ... & P030402406003 & 1284 \\ [6pt]
        \bottomrule
    \end{tabular}
\end{table*}

\begin{figure*}
     \subfloat{
         \includegraphics[width=0.49\textwidth]{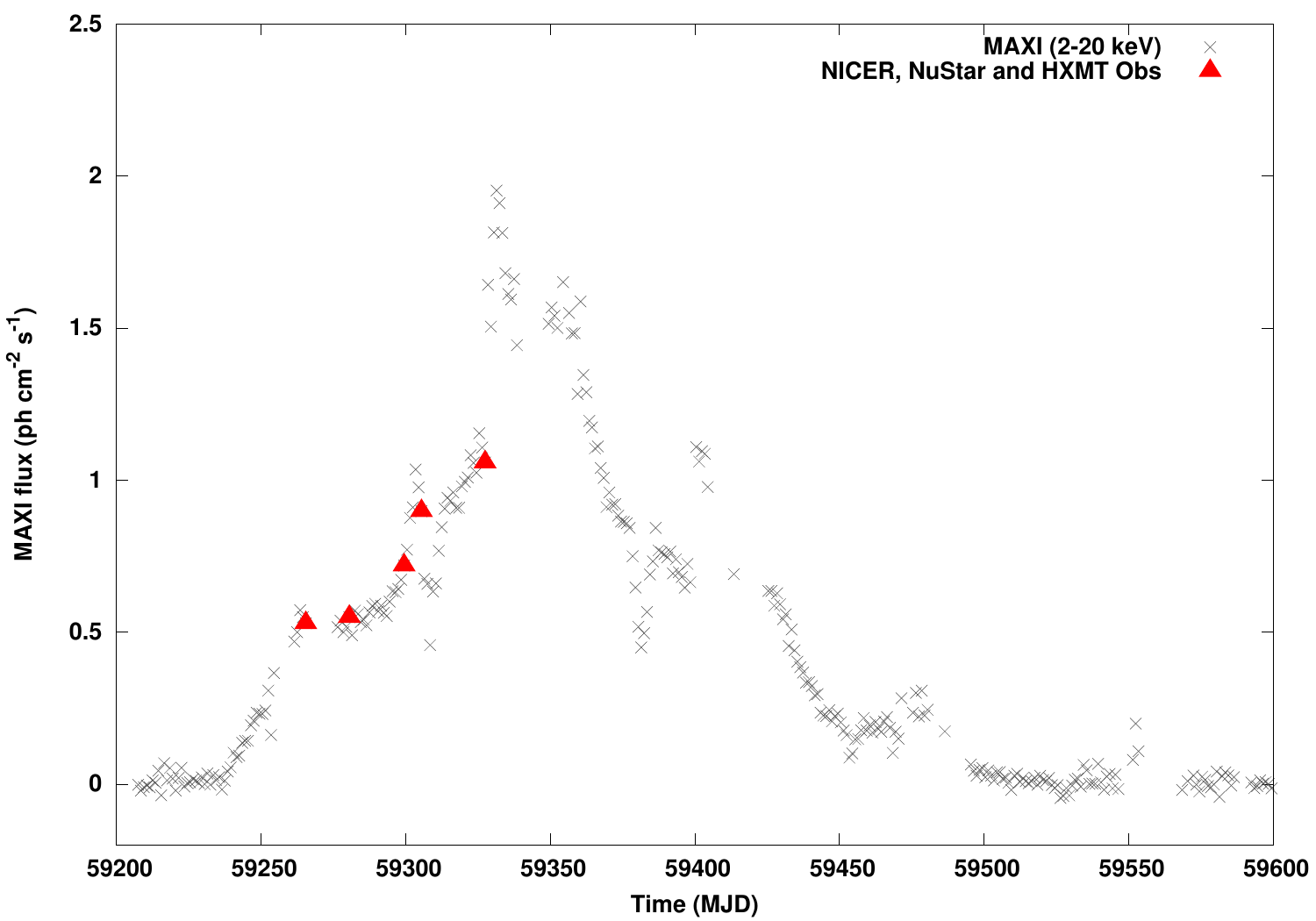}
     }
     \hfill
     \subfloat{
         \includegraphics[width=0.49\textwidth]{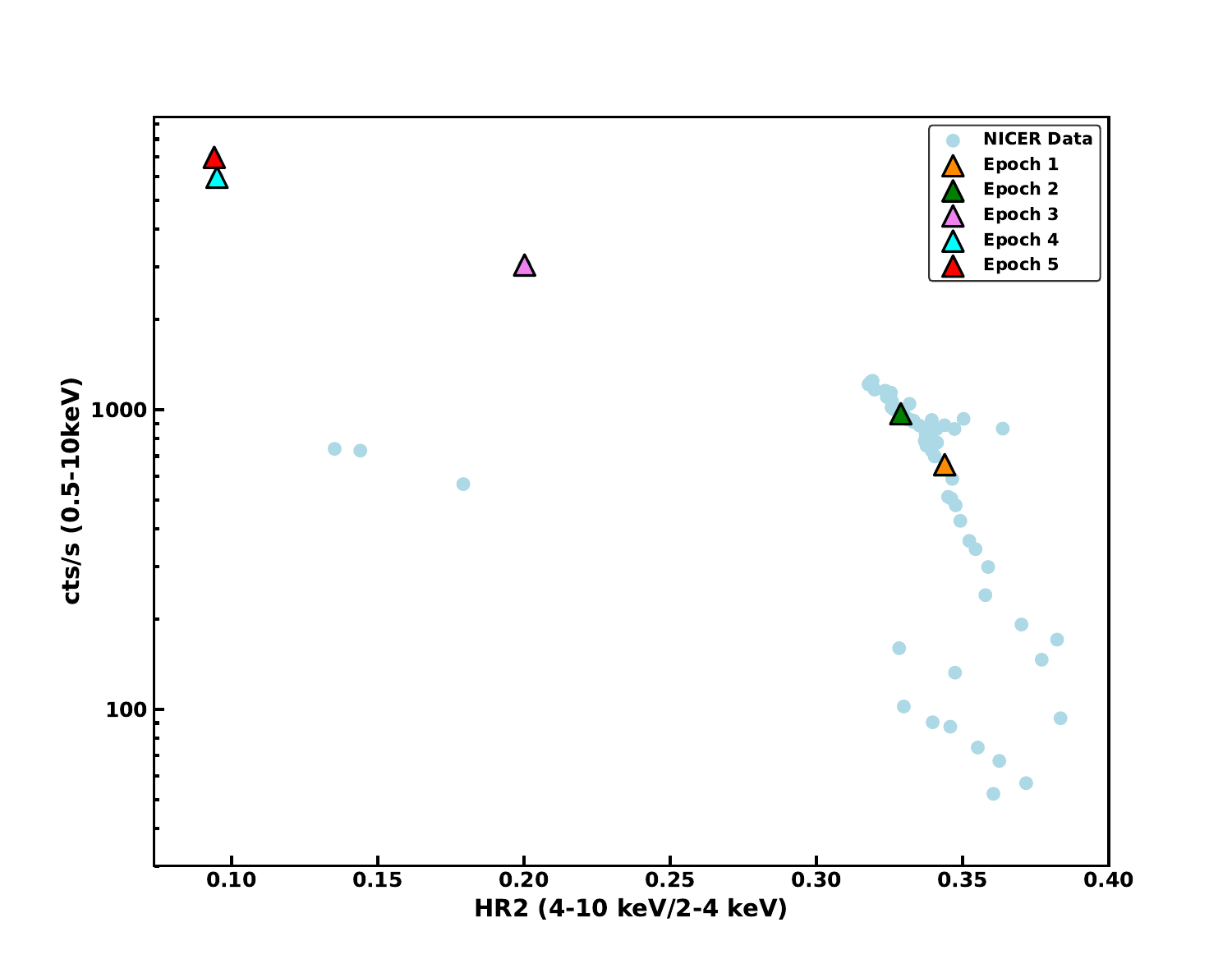}
     }
     \caption{The left panel displays the one-day binned light curve of GX 339-4, derived from MAXI data. The red triangles indicate the specific observations utilized in this analysis, with the corresponding dates detailed in Table \ref{tab:table1}. The right panel presents the Hardness–Intensity Diagram (HID) for GX 339-4, plotted using NICER data. The diagram shows the evolution of the hardness ratio (4-10 keV/2–4 keV) on the x-axis and the count rate in the 0.5–10 keV range on the y-axis. It highlights the simultaneous NICER, NuSTAR, and HXMT observations that are included in this study. }
\label{fig:fig1}
\end{figure*}

%%%%%%%%%%%%%%%%%%%%%%%%%%%%%%%%%%%%%%%%%%%%%%%%%%%%%%%%%%%%%%%%%%%%%%%%%%%%%%%%%%%%%
\section{Observations and Data Reduction}
\label{sec:obs_data_red}
In 2021, X-ray monitoring detected a new outburst of GX 339–4 and triggered observations with the NuSTAR, NICER, and HXMT. We used all the simultaneous observations in these three satellites to study the transition of the source GX339-4 from hard to soft state.

Our primary objective is to investigate the transition of GX 339-4 to its soft state. This study utilizes observations made on 20 Feb 2021 (Epoch 1), 8 Mar 2021 (Epoch 2), 26 Mar 2021 (Epoch 3), 1 Apr 2021 (Epoch 4), and 24 Apr 2021 (Epoch 5). Table \ref{tab:table1} lists the observation IDs and effective exposure times for NICER, NuSTAR, and HXMT data corresponding to each epoch. To track the source's evolution, we extracted the MAXI 2-20 keV flux, as shown in the left panel of Fig. \ref{fig:fig1}. This light curve spans from 17 Dec 2020 to 21 Jan 2022, with red triangles marking the observations used in this study. The triangle sequence in the light curve matches the order in Table \ref{tab:table1}. The right panel of \ref{fig:fig1} illustrates the Hardness–Intensity Diagram (HID) for GX 339-4, obtained using NICER data. In this diagram, the x-axis represents the evolution of the hardness ratio (4–10 keV/2–4 keV), while the y-axis shows the count rate in the 0.5–10 keV energy band. The HID highlights the epochs with simultaneous NICER, NuSTAR, and HXMT observations used in this analysis. The source transitioned from hard to soft states during these observations. We aim to investigate how the system's properties change during this transition.

\subsection{HXMT Data Reduction} 
\textit{Insight}-HXMT \citep[][]{2020SCPMA..6349502Z} comprises three primary instruments: the High Energy X-ray Telescope (HE, 20–250 keV; \citet{2020SCPMA..6349503L}), the Medium Energy X-ray Telescope (ME, 5–40 keV; \citet{2020SCPMA..6349504C}), and the Low Energy X-ray Telescope (LE, 1–12 keV; \citet{2020SCPMA..6349505C}). The data reduction was performed according to the HXMT Data Reduction Guide v2.04\footnote{\label{ft4}\href{http://hxmtweb.ihep.ac.cn/SoftDoc/496.jhtml}{http://hxmtweb.ihep.ac.cn/SoftDoc/496.jhtml}}, using the {\fontfamily{pcr}\selectfont HXMTDAS} v2.04 software. Background spectra are generated using the \texttt{hebkgmap}, \texttt{mebkgmap}, and \texttt{lebkgmap} scripts, as described in \citet{2020JHEAp..27...44G}, \citet{2020JHEAp..27...14L} and \citet{2020JHEAp..27...24L}.
We combined the HXMT spectra that were simultaneous with NICER and NuSTAR observations using the {\fontfamily{pcr}\selectfont addspec} tool from the {\fontfamily{pcr}\selectfont ftools} package in XSPEC version 12.12, and merged the simultaneous light curves with the {\fontfamily{pcr}\selectfont ftmerge} tool.  For the analysis, we utilized the 2–10 keV band for LE, 10–30 keV for ME, and 30–100 keV for HE. Barycentric corrections were applied to the light curves using the {\fontfamily{pcr}\selectfont hxbary} tool. A systematic uncertainty\footnote{\href{http://hxmten.ihep.ac.cn/mission.jhtml}{http://hxmten.ihep.ac.cn/mission.jhtml}} of 2\% was applied to LE and ME spectra, while a 1\% systematic uncertainty was applied to HE spectra.

\begin{figure*}
      \vspace{-10pt} % reduce space above the figure
     \subfloat{
         \includegraphics[width=0.49\textwidth]{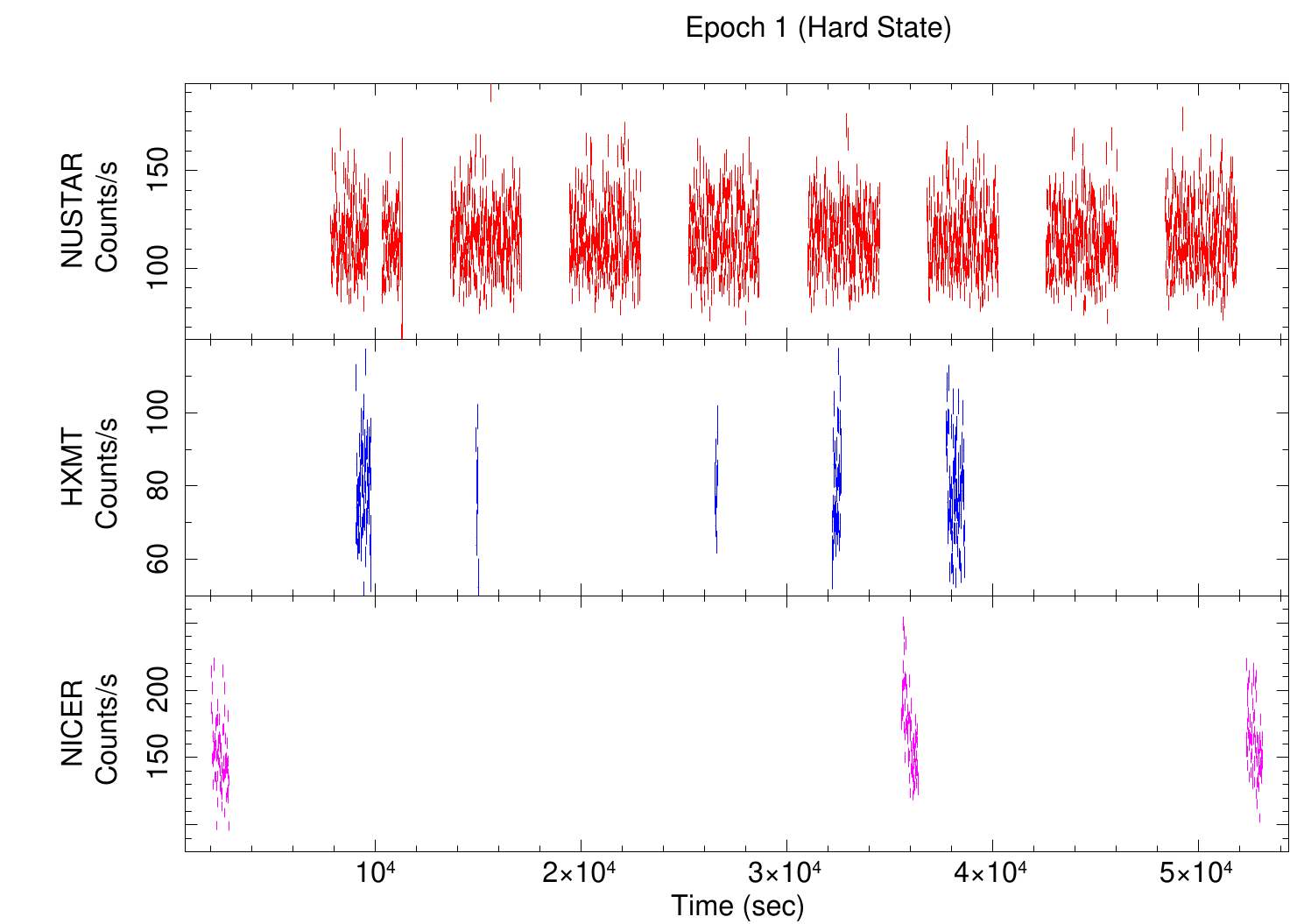}
     }
     \hfill
     \subfloat{
         \includegraphics[width=0.49\textwidth]{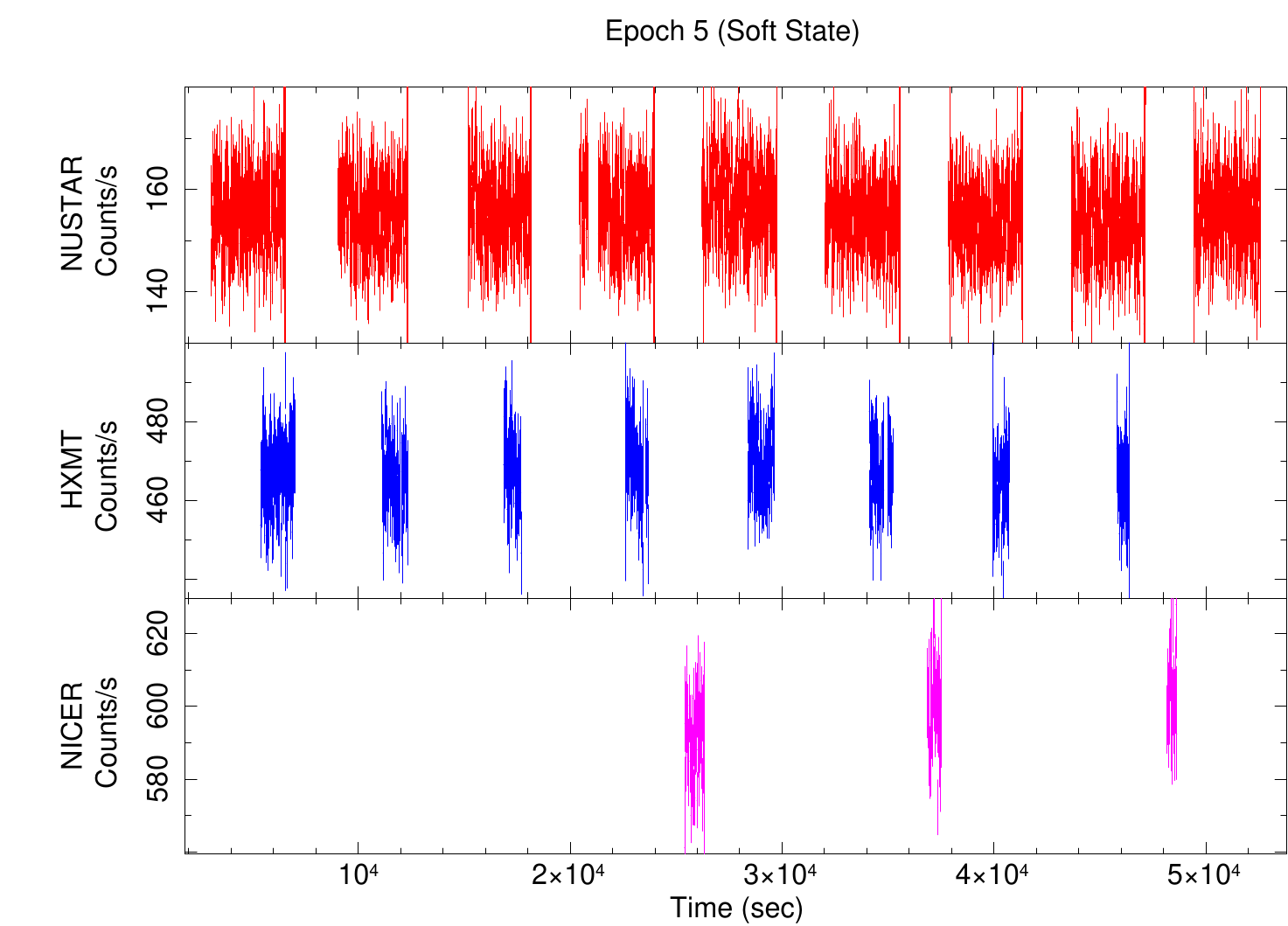}
     }
     \caption{The 10-second binned, barycentric-corrected light curves from NuSTAR (red, 3-80 keV), HXMT (LE, blue, 1-10 keV), and NICER (magenta, 2-12 keV) during two distinct states of the source. Left: Hard state (Epoch 1). Right: Soft state (Epoch 5).}

\label{fig:fig2}
\end{figure*}

\subsection{NICER Data Reduction}
The Neutron Star Interior Composition Explorer (NICER) \citep[][]{2016SPIE.9905E..1HG,2012SPIE.8443E..13G} is an instrument onboard the International Space Station (ISS). The primary instrument of NICER is the X-ray Timing Instrument
(XTI), which is an array of 56 photon detectors that operate in the 0.2–12 keV energy range. \citep[][]{2016SPIE.9905E..1HG}. Currently, 52 detectors are operational. NICER observed the GX 339-4 source multiple times during its 2021 outburst.

\begin{table*}
\setlength{\tabcolsep}{10pt}
\centering
\caption{Broad-band X-ray spectral parameters of GX~339--4 from simultaneous NICER, NuSTAR, and Insight-HXMT observations, analysed using both single and dual Comptonization models.}
\label{tab:table2}
\begin{tabular}{c c c c c}
\hline \hline
Components & Parameters &
Epoch 1 (Single thcomp) &
Epoch 1 (Dual thcomp) &
Epoch 5 (Single thcomp) \\ [2pt]
\hline
Tbabs
& $N_H$ ($10^{22}\,\mathrm{cm^{-2}}$)
& $0.23^{+0.02}_{-0.01}$
& $0.39^{+0.05}_{-0.06}$
& $0.48^{+0.01}_{-0.01}$ \\ [2pt]

Hard Comptonization
& $\Gamma$
& $1.64^{+0.01}_{-0.01}$
& $1.62^{+0.01}_{-0.01}$
& $2.13^{+0.01}_{-0.01}$ \\ [2pt]

& $kT_e$ (keV)
& $25.6^{+1.9}_{-0.6}$
& $23.5^{+1.0}_{-1.0}$
& $>140$ \\ [2pt]

& $f_{\mathrm{SC}}$
& $0.78^{+0.02}_{-0.01}$
& $0.38^{+0.06}_{-0.07}$
& $0.011^{+0.001}_{-0.001}$ \\ [2pt]

Soft Comptonization
& $\Gamma$
& --
& $2.21^{+0.06}_{-0.07}$
& -- \\ [2pt]

& $kT_e$ (keV)
& --
& $1.6^{+0.3}_{-0.2}$
& -- \\ [2pt]

& $f_{\mathrm{SC}}$
& --
& $>0.35$
& -- \\ [2pt]

Disk
& $T_{\mathrm{in}}$ (keV)
& $0.78^{+0.03}_{-0.03}$
& $0.20^{+0.12}_{-0.13}$
& $0.81^{+0.01}_{-0.01}$ \\ [2pt]

& Norm. ($10^{3}$)
& $0.28^{+0.05}_{-0.03}$
& $>2.6$
& $3.02^{+0.03}_{-0.04}$ \\ [2pt]

Reflection
& Incl. (deg)
& $28.5^{+3.3}_{-1.0}$
& $19.6^{+4.0}_{-4.9}$
& $19.4^{+1.7}_{-2.2}$ \\ [2pt]

& $R_{\mathrm{in}}$ ($r_g$)
& $5.3^{+1.1}_{-1.1}$
& $14.3^{+2.4}_{-2.8}$
& $<2.5$ \\ [2pt]

& $\log \xi$ ($\mathrm{erg\,cm\,s^{-1}}$)
& $3.52^{+0.05}_{-0.11}$
& $3.45^{+0.08}_{-0.20}$
& $3.70^{+0.04}_{-0.05}$ \\ [2pt]

& $A_{\mathrm{Fe}}$
& $>7.0$
& $2.9^{+2.0}_{-2.2}$
& $>9.3$ \\ [2pt]

& Norm. ($10^{-3}$)
& $2.3^{+0.5}_{-0.1}$
& $1.7^{+0.4}_{-0.2}$
& $2.8^{+0.1}_{-0.1}$ \\ [2pt]

\hline
& $\chi^2/\mathrm{bins}$ (FPMA)
& 290.8/264 & 278.6/264 & 253.5/214 \\ [2pt]
& $\chi^2/\mathrm{bins}$ (FPMB)
& 275.7/268 & 253.7/268 & 267.1/218 \\ [2pt]
& $\chi^2/\mathrm{bins}$ (LE)
& 62.3/50 & 50.6/50 & 61.2/54 \\ [2pt]
& $\chi^2/\mathrm{bins}$ (ME)
& 12.3/14 & 13.4/14 & 7.1/14 \\ [2pt]
& $\chi^2/\mathrm{bins}$ (HE)
& 4.4/14 & 6.4/14 & 9.0/15 \\ [2pt]
& $\chi^2/\mathrm{bins}$ (XTI)
& 121.5/142 & 86.6/142 & 79.4/143 \\ [2pt]
& $\chi^2/\mathrm{dof}$ (Total)
& 766.5/736 & 689.3/733 & 677.4/642 \\ [2pt]
\hline \hline
\end{tabular}
\end{table*}

We processed the NICER data using the pipeline tool {\fontfamily{pcr}\selectfont nicerl2}\footnote{\href{https://heasarc.gsfc.nasa.gov/lheasoft/ftools/headas/nicerl2.html}{https://heasarc.gsfc.nasa.gov/lheasoft/ftools/headas/nicerl2.html}} available in NICERDAS v11 with HEASOFT v6.32 and applied standard filters. The calibration database version used was xti20240206.
Data from detectors 14 and 34, which were affected by increased electronic noise, were excluded using the HEASOFT routine {\fontfamily{pcr}\selectfont fselect}. Barycenter corrections for each NICER observation were applied using the FTOOL BARYCORR. The spectra, along with the corresponding response and area files, were generated using {\fontfamily{pcr}\selectfont nicerl3-spect}, which also applied a systematic error of approximately 1.5\% using {\fontfamily{pcr}\selectfont niphasyserr}. Background spectra were produced using the Scorpeon model in {\fontfamily{pcr}\selectfont nicerl3}. Light curves in the 2-12 keV energy range were extracted using {\fontfamily{pcr}\selectfont nicerl3-lc}.  We utilized NICER data in 1-10 keV for spectral analysis.

\subsection{NuSTAR Data Reduction}
NuSTAR is the first mission to employ focusing telescopes for high-energy X-ray (3–79 keV) sky imaging within the electromagnetic spectrum. It was launched on 2012 June 13, at 9 am PDT \citep[][]{2013ApJ...770..103H}. For NuSTAR data analysis, we employ NuSTARDAS v2.1.2 software with the calibration database (CALDB 20230613). We extract NuSTAR-filtered data using the standard pipeline program {\fontfamily{pcr}\selectfont nupipeline}. 
Source spectra were extracted from a 120-arcsecond circular region centered on GX 339-4, with background spectra obtained from a similarly sized circular region away from the source. For spectral analysis, we utilized data from both FPMA and FPMB detectors in the 3-79 keV energy range. \\

\section{Results}
\label{sec:result}
Spectral analysis was performed using XSPEC v12.11.1 \citep[][]{1996ASPC..101...17A}. 
The extracted spectra were grouped using the optimal binning scheme (via \texttt{ftgrouppha} tool from FTOOLS\footnote{\href{https://heasarc.gsfc.nasa.gov/lheasoft/ftools/headas/ftgrouppha.html}{https://heasarc.gsfc.nasa.gov/lheasoft/ftools/headas/ftgrouppha.html}} as implemented in the NICER spectral pipeline), which applies finer binning where the spectrum/response contains sharper features and coarser binning where the statistical quality is lower, thereby avoiding unnecessary oversampling of the instrumental resolution \citep{KaastraBleeker2016}. Parameter estimation was carried out using the $\chi^2$ statistic, with uncertainties quoted at the 90\% confidence level. To account for interstellar absorption along the line of sight, we included the {\fontfamily{pcr}\selectfont TBabs} component \citep[][]{2000ApJ...542..914W}, adopting the abundance set from \citet{1989GeCoA..53..197A} and photoelectric cross-sections from \citet{1996ApJ...465..487V}. The hydrogen column density ($N_{\rm H}$) was left free to vary across all epochs. A cross-normalization constant was included to account for calibration differences among the instruments.

\begin{figure*}
\vspace{-10pt} % reduce space above the figure
     \subfloat{
         \includegraphics[width=0.49\textwidth]{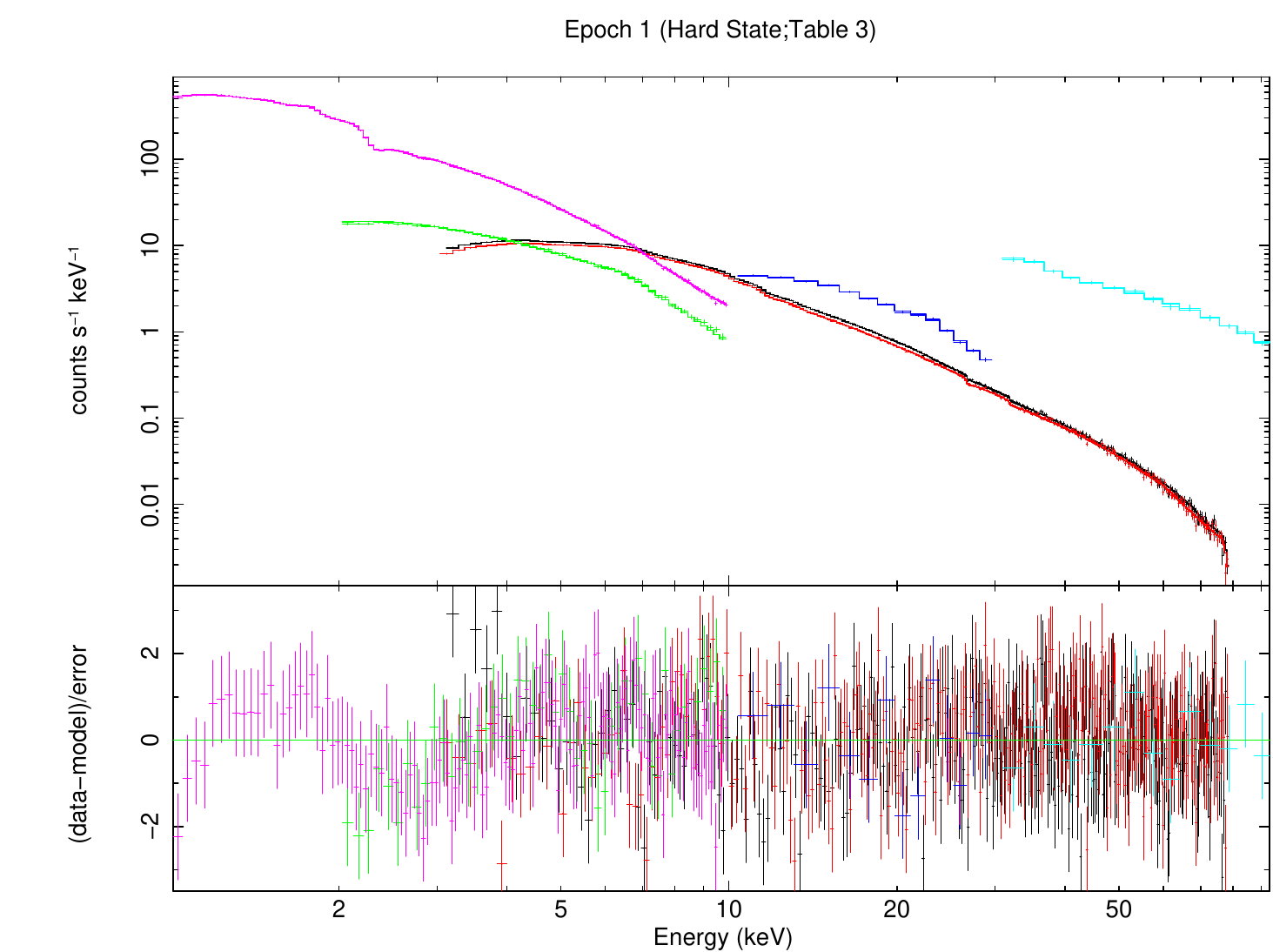}
     }
     \hfill
     \subfloat{
         \includegraphics[width=0.49\textwidth]{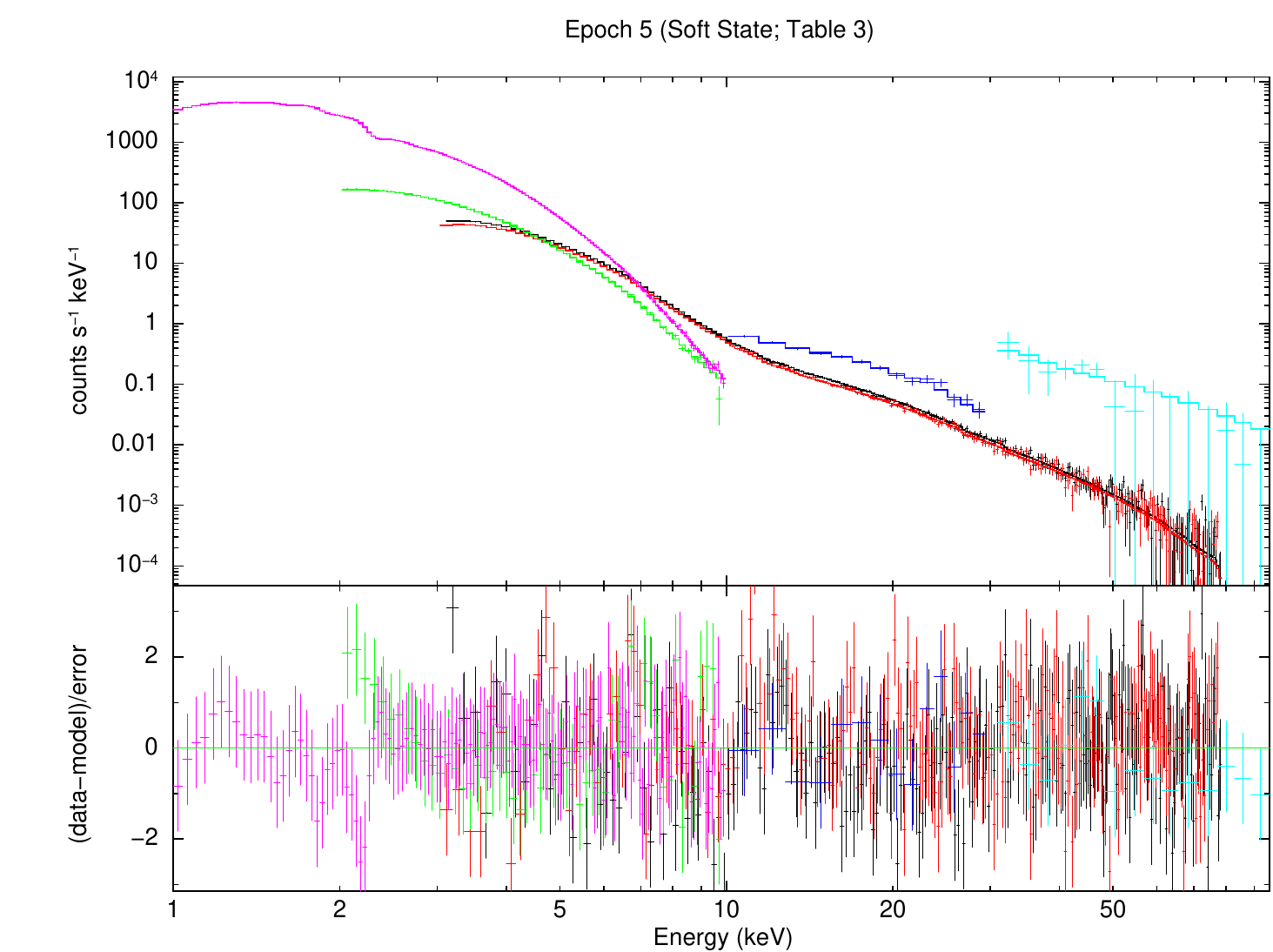}
     }
     
     \caption{Simultaneous broadband spectra of GX~339$-$4 in the hard state (left; Epoch~1) and soft state (right; Epoch~5), fitted with the single-Comptonization model (Table~3). Data are shown from NICER (pink), NuSTAR/FPMA (red), NuSTAR/FPMB (black), \textit{Insight}-HXMT/LE (green), \textit{Insight}-HXMT/ME (blue), and \textit{Insight}-HXMT/HE (cyan). The lower panels show the residuals, $(\mathrm{data}-\mathrm{model})/\sigma$, for the best-fitting models.}

\label{fig:fig3}
\end{figure*}

\begin{table*}
\setlength{\tabcolsep}{13pt}
\centering
\caption{Broad-band X-ray spectral parameters of GX~339--4 derived from simultaneous \textit{NICER}, \textit{NuSTAR}, and \textit{Insight--HXMT} observations, modeled using a single Comptonization component across all epochs. Best-fit parameters correspond to the model \texttt{tbabs*(thcomp*diskbb + relxillCp)}. The inclination angle and iron abundance ($A_\mathrm{Fe}$) are fixed at their average values of $30^\circ$ and 9 times the solar abundance, respectively. See text for further details.}
\label{tab:table3}
%\begin{tabular}{ M{2.7cm} M{1.7cm} M{2.4cm} M{2.4cm} M{2.3cm} M{2.cm} }
\begin{tabular}{c c c c c c c} 
\hline \hline
\textbf{Components} & \textbf{Parameters} & \textbf{Epoch 1}  & \textbf{Epoch 2} & \textbf{Epoch 3} & \textbf{Epoch 4} & \textbf{Epoch 5} \\
\hline
      Tbabs          &  $N_H$ ($10^{22} \textrm{cm}^{-2}$)  & $0.26^{+0.01}_{-0.02}$      & $0.27^{+0.01}_{-0.01}$        & $0.43^{+0.01}_{-0.01}$        & $0.46^{+0.01}_{-0.01}$         & $0.47^{+0.01}_{-0.01}$        \\  [2pt]
  Comptonization     &   $\Gamma$            & $1.64^{+0.01}_{-0.01}$      & $1.64^{+0.01}_{-0.01}$        & $2.03^{+0.01}_{-0.01}$        & $2.20^{+0.02}_{-0.01}$         & $2.15^{+0.03}_{-0.03}$            \\  [2pt]
& $kT_e$ (keV)        & $44.9^{+2.8}_{-3.0}$      & $26.8^{+1.0}_{-0.9}$        & $87^{+47}_{-21}$      & $>180$   & $68^{+78}_{-24}$       \\[2pt]
 & $f_{SC}$             & $0.87^{+0.02}_{-0.02}$    & $0.76^{+0.01}_{-0.01}$      & $0.37^{+0.02}_{-0.03}$      & $0.051^{+0.004}_{-0.005}$       & $0.012^{+0.003}_{-0.002}$        \\[2pt]
      Disk           &  $T_{in}$ (keV)               & $0.71^{+0.03}_{-0.03}$      & $0.70^{+0.02}_{-0.02}$        & $0.55^{+0.01}_{-0.01}$        & $0.77^{+0.01}_{-0.01}$         & $0.81^{+0.01}_{-0.01}$           \\  [2pt]
                     &   Norm ($10^{3}$)     & $0.38^{+0.05}_{-0.06}$      & $0.43^{+0.05}_{-0.04}$     & $4.08^{+0.19}_{-0.21}$     & $2.95^{+0.04}_{-0.05}$         & $2.92^{+0.03}_{-0.03}$           \\  [2pt]
  Reflection         &   $R_{in}$($r_g$)     & $4.78^{+0.57}_{-0.33}$      & $4.77^{+0.33}_{-0.14}$        & $1.96^{+0.22}_{-1.96}$        & $2.56^{+0.24}_{-0.26}$         & $<2.1$         \\  [2pt]
                     &   $\log\xi \ (\mathrm{erg \ cm \ s^{-1}})$     & $3.47^{+0.03}_{-0.04}$      & $3.50^{+0.04}_{-0.04}$        & $4.03^{+0.06}_{-0.04}$        & $3.80^{+0.05}_{-0.05}$         & $3.32^{+0.09}_{-0.07}$           \\  [2pt]
& Norm ($10^{-3}$)    & $3.05^{+0.12}_{-0.14}$       & $2.55^{+0.07}_{-0.06}$      & $9.6^{+0.8}_{-0.9}$      & $4.41^{+0.21}_{-0.21}$          & $2.78^{+0.29}_{-0.20}$           \\[2pt]

\hline
& $\chi^2/\mathrm{bins}$ (FPMA)    
& 291.6/264     
& 296.0/264     
& 367.9/245     
& 223.0/220     
& 276.4/214   \\[2pt]

& $\chi^2/\mathrm{bins}$ (FPMB)     
& 277.0/268     
& 280.3/268     
& 328.2/247     
& 343.6/224     
& 298.1/218   \\[2pt]

& $\chi^2/\mathrm{bins}$ (LE)       
& 62.3/50     
& 46.9/48     
& 50.8/52     
& 52.4/51     
& 63.4/54             \\[2pt]

& $\chi^2/\mathrm{bins}$ (ME)    
& 12.4/14     
& 28.3/14     
& 5.4/14     
& 13.8/14     
& 6.7/14               \\[2pt]

& $\chi^2/\mathrm{bins}$ (HE)      
& 4.3/14     
& 3.8/15     
& 7.7/15     
& 4.2/14     
& 8.6/15                  \\[2pt]

& $\chi^2/\mathrm{dof}$ (Total)  
& 766.9/738     
& 769.0/746     
& 826.0/709     
& 807.1/663     
& 728.9/644   \\[2pt]

\hline \hline
\end{tabular}
\end{table*}

\begin{figure*}
\vspace{-10pt} % reduce space above the figure
     \subfloat{
         \includegraphics[width=0.49\textwidth]{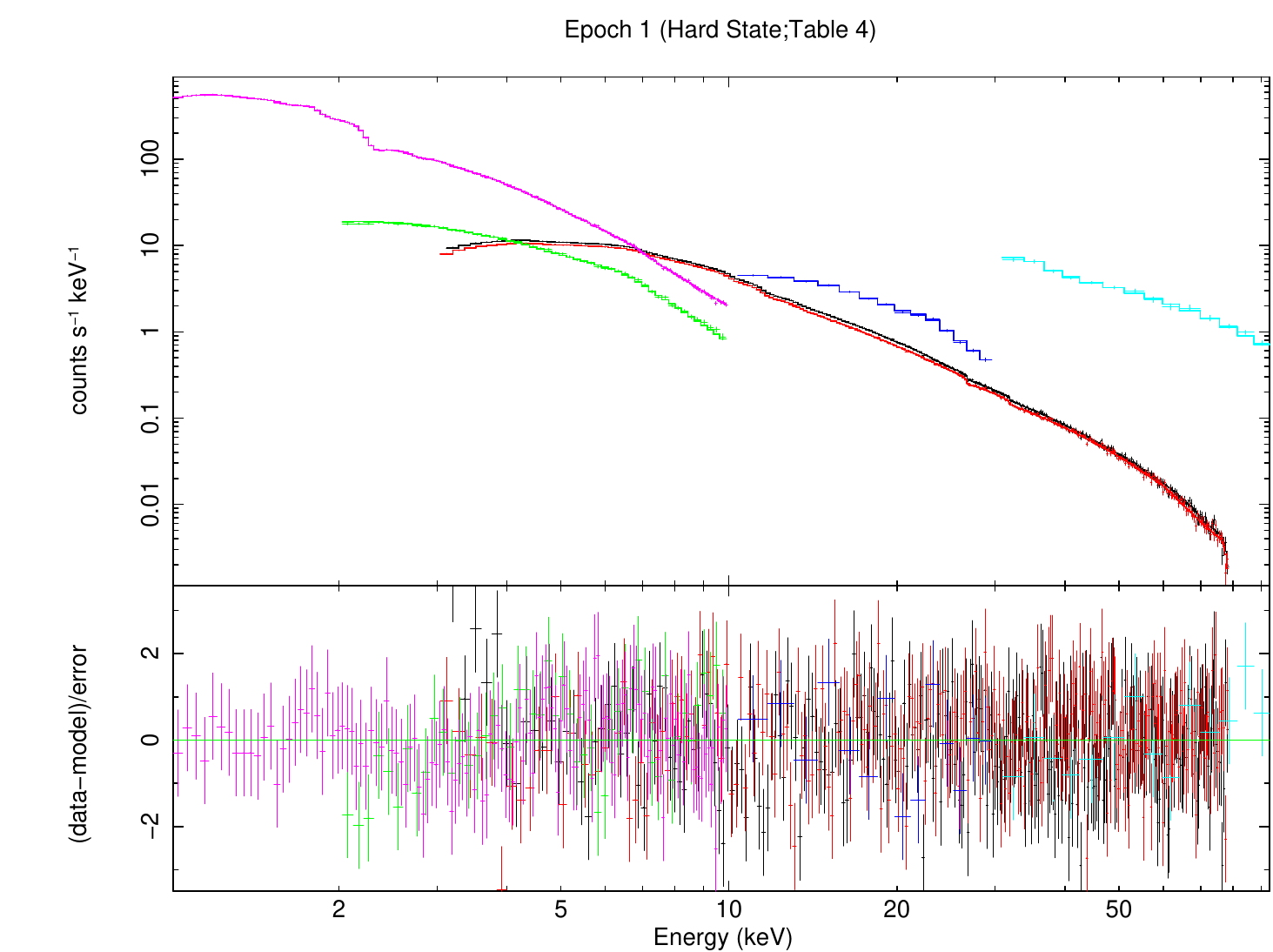}
     }
     \hfill
     \subfloat{
         \includegraphics[width=0.49\textwidth]{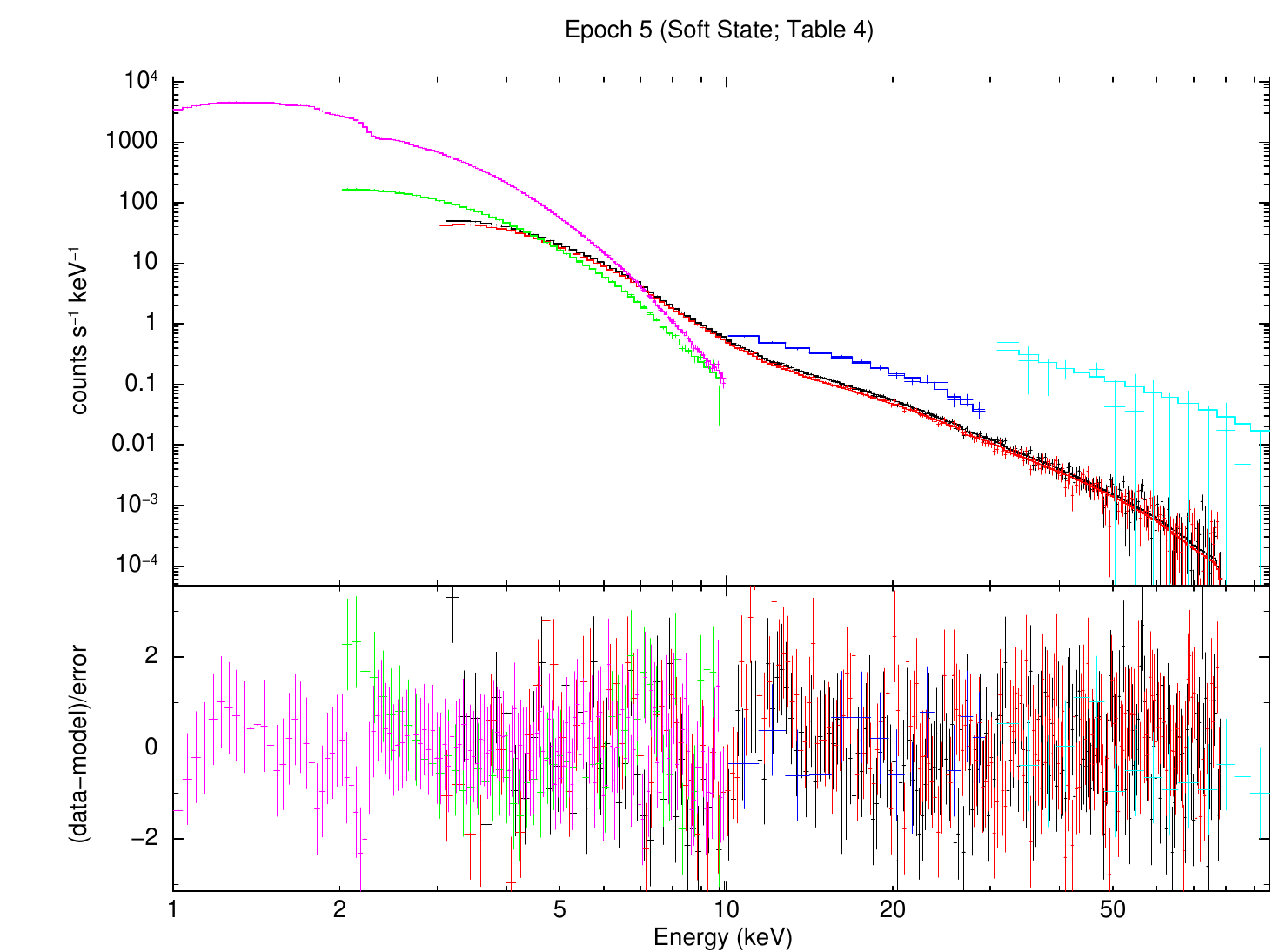}
     }
     
     \caption{Simultaneous broadband spectra of GX~339$-$4 in the hard state (left; Epoch 1) and soft state (right; Epoch 5), fitted with the best-fitting models reported in Table~4. Data are shown from NICER (pink), NuSTAR/FPMA (red), NuSTAR/FPMB (black), \textit{Insight}-HXMT/LE (green), \textit{Insight}-HXMT/ME (blue), and \textit{Insight}-HXMT/HE (cyan). The lower panels show the residuals, for the best-fitting models.}

\label{fig:fig4}
\end{figure*}

\begin{table*}
\setlength{\tabcolsep}{8pt}
\centering
\caption{Broad-band X-ray spectral parameters of GX~339--4 derived from simultaneous \textit{NICER}, \textit{NuSTAR}, and \textit{Insight--HXMT} observations, modeled using a dual Comptonization component for all epochs. Best-fit parameters correspond to the model \texttt{tbabs*(thcomp*thcomp*diskbb + relxillCp)}. The inclination angle and iron abundance ($A_\mathrm{Fe}$) are fixed at their average values of $22.6^\circ$ and 8.3 times the solar abundance, respectively. The unabsorbed flux is calculated in the 0.1--100~keV energy band using the \texttt{cflux} model in \texttt{XSPEC}.}
\label{tab:table4}
\begin{tabular}{c c c c c c c} 
\hline \hline

\textbf{Components} & \textbf{Parameters} & \textbf{Epoch 1}  & \textbf{Epoch 2} & \textbf{Epoch 3} & \textbf{Epoch 4} & \textbf{Epoch 5} \\
\hline   \\
Tbabs & $N_H$ ($10^{22}\,\textrm{cm}^{-2}$) & $0.33^{+0.05}_{-0.03}$ & $0.42^{+0.06}_{-0.05}$ & $0.44^{+0.02}_{-0.02}$ & $0.49^{+0.01}_{-0.01}$ & $0.48^{+0.01}_{-0.01}$ \\[2pt]

Hard Comptonization & $\Gamma$ & $1.62^{+0.01}_{-0.01}$ & $1.62^{+0.01}_{-0.01}$ & $1.94^{+0.02}_{-0.02}$ & $2.13^{+0.03}_{-0.02}$ & $2.14^{+0.01}_{-0.02}$ \\[2pt]

& $kT_e$ (keV) & $22.9^{+0.9}_{-0.8}$ & $23.1^{+0.9}_{-0.9}$ & $28.6^{+3.7}_{-2.8}$ & $37.6^{+12}_{-8}$ & $>180$ \\[2pt]

& $f_{SC}$ & $0.54^{+0.05}_{-0.10}$ & $0.38^{+0.08}_{-0.12}$ & $0.19^{+0.02}_{-0.02}$ & $0.037^{+0.004}_{-0.003}$ & $0.011^{+0.001}_{-0.002}$ \\[2pt]

Soft Comptonization & $\Gamma$ & $2.19^{+0.08}_{-0.12}$ & $2.46^{+0.07}_{-0.06}$ & $2.2^{+0.9}_{-1.2}$ & $2.8^{+3.2}_{-0.5}$ & -- \\[2pt]

& $kT_e$ (keV) & $1.37^{+0.15}_{-0.15}$ & $1.73^{+0.28}_{-0.23}$ & $1.3^{+0.7}_{-0.2}$ & $0.90^{+0.06}_{-0.90}$ & -- \\[2pt]

& $f_{SC}$ & $>0.45$ & $>0.43$ & $0.19^{+0.30}_{-0.18}$ & $>0.51$ & -- \\[2pt]

Disk & $T_{in}$ (keV) & $0.27^{+0.15}_{-0.07}$ & $0.20^{+0.21}_{-0.06}$ & $0.48^{+0.05}_{-0.04}$ & $0.62^{+0.02}_{-0.04}$ & $0.81^{+0.01}_{-0.03}$ \\[2pt]

& Norm. ($10^{3}$) & $>10.1$ & $>20.7$ & $8.3^{+2.8}_{-2.0}$ & $6.3^{+0.6}_{-0.9}$ & $2.99^{+0.04}_{-0.04}$ \\[2pt]

Reflection & $R_{in}$ ($r_g$) & $10.4^{+1.2}_{-1.0}$ & $9.6^{+1.0}_{-0.8}$ & $8.1^{+0.9}_{-0.8}$ & $8.3^{+1.3}_{-1.2}$ & $<2.7$ \\[2pt]

& $\log\xi$ ($\mathrm{erg\,cm\,s^{-1}}$) & $3.48^{+0.04}_{-0.05}$ & $3.51^{+0.04}_{-0.04}$ & $4.12^{+0.06}_{-0.08}$ & $3.52^{+0.11}_{-0.10}$ & $3.61^{+0.07}_{-0.07}$ \\[2pt]

& Norm ($10^{-3}$) & $1.63^{+0.09}_{-0.09}$ & $1.74^{+0.07}_{-0.07}$ & $4.8^{+0.8}_{-0.8}$ & $1.56^{+0.22}_{-0.09}$ & $3.06^{+0.14}_{-0.39}$ \\[2pt]

cflux & $F_{\mathrm{total}}$ ($10^{-8}\,\mathrm{erg\,cm^{-2}\,s^{-1}}$) & $1.43^{+0.09}_{-0.58}$ & $1.51^{+0.14}_{-0.58}$ & $2.22^{+0.04}_{-0.07}$ & $2.77^{+0.04}_{-0.05}$ & $3.09^{+0.03}_{-0.03}$ \\[2pt]

& $F_{\mathrm{diskbb}}$ ($10^{-8}\,\mathrm{erg\,cm^{-2}\,s^{-1}}$) & $0.14^{+0.03}_{-0.26}$ & $0.18^{+0.05}_{-0.33}$ & $0.91^{+0.03}_{-0.03}$ & $2.00^{+0.10}_{-0.40}$ & $2.73^{+0.01}_{-0.01}$ \\[2pt]

\hline
& $\chi^2/\mathrm{bins}$ (FPMA) & 285.3/264 & 281.4/264 & 289.7/245 & 202.0/220 & 259.2/214 \\[2pt]
& $\chi^2/\mathrm{bins}$ (FPMB) & 256.4/268 & 252.7/268 & 291.0/247 & 275.3/224 & 272.3/218 \\[2pt]
& $\chi^2/\mathrm{bins}$ (LE)   & 51.2/50   & 48.0/48   & 54.3/52   & 40.7/51   & 61.9/54  \\[2pt]
& $\chi^2/\mathrm{bins}$ (ME)   & 12.8/14   & 27.0/14   & 7.4/14    & 13.8/14   & 7.0/14   \\[2pt]
& $\chi^2/\mathrm{bins}$ (HE)   & 7.8/14    & 9.0/15    & 9.4/15    & 4.8/14    & 8.8/15   \\[2pt]
& $\chi^2/\mathrm{bins}$ (XTI)  & 85.9/142  & 74.6/151  & 56.4/150  & 142.0/154 & 77.6/143 \\[2pt]
& $\chi^2/\mathrm{dof}$ (Total) & 699.4/735 & 692.8/743 & 708.0/706 & 678.6/660 & 686.8/644 \\[2pt]
\hline
\end{tabular}
\end{table*}

The X-ray spectra of GX 339–4 across different accretion states are well described by a combination of thermal emission from the accretion disk and nonthermal Comptonized emission from a hot corona, consistent with the typical phenomenology of BHXBs \citep[][]{1995ApJ...445..780S, 2004PThPS.155...99Z}. To model the thermal component, we used the multicolor disk blackbody model {\fontfamily{pcr}\selectfont Diskbb} \citep[][]{1984PASJ...36..741M, 1986ApJ...308..635M}, characterized by the inner disk temperature ($T_{\rm in}$) and normalization ($N_{\rm Disk}$). The \texttt{diskbb} normalization ($N_{\mathrm{DBB}}$) is related to the apparent inner disc radius $r_{\mathrm{in}}$ through $N_{\mathrm{DBB}} = \left( \frac{r_{\mathrm{in}}}{D_{10}} \right)^2 \cos\theta$, where $r_{\mathrm{in}}$ is in km, $D_{10}$ is the source distance in units of 10 kpc, and $\theta$ is the disc inclination angle. The physical inner disc radius is given by $R_{\mathrm{in}} = \xi \kappa^2 r_{\mathrm{in}}$, where $\xi = 0.41$ accounts for the stress-free inner boundary condition \citep[][]{Kubota1998}, and $\kappa = 1.7$--$2.0$ is the spectral hardening factor \citep{1995ApJ...445..780S}. The high-energy nonthermal tail was modeled using the physically motivated Comptonization model {\fontfamily{pcr}\selectfont thcomp} \citep[][]{2020MNRAS.492.5234Z}, which includes parameters for the photon index ($\Gamma$), electron temperature ($kT_{\rm e}$), and scattering fraction ($f_{\rm sc}$) (e.g., \citet{2025ApJ...986...97C}, \citet{2024ApJ...974...90D}).

\begin{figure*}
\vspace{-10pt} % reduce space above the figure
     \subfloat{
         \includegraphics[width=0.49\textwidth]{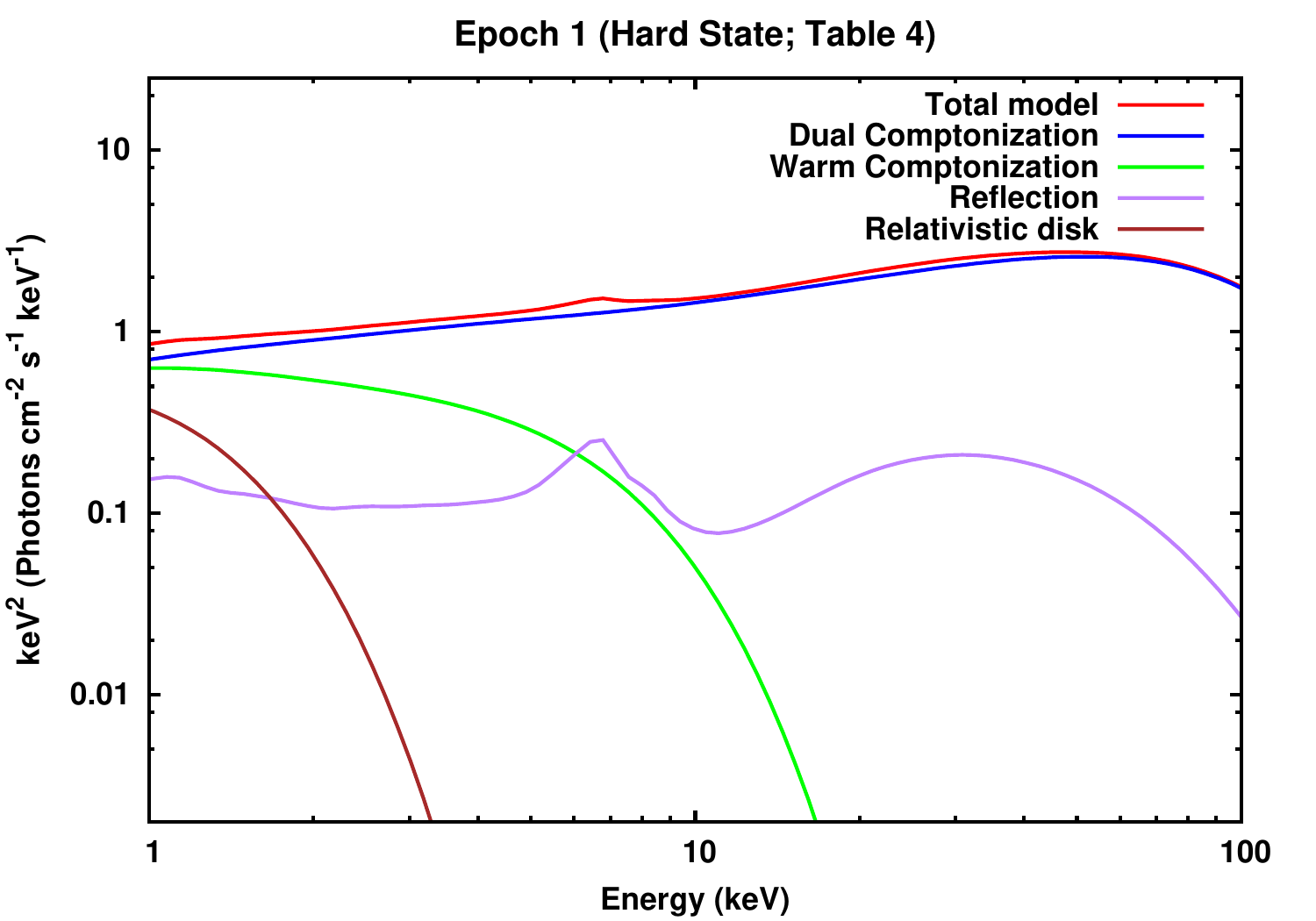}
     }
     \hfill
     \subfloat{
         \includegraphics[width=0.49\textwidth]{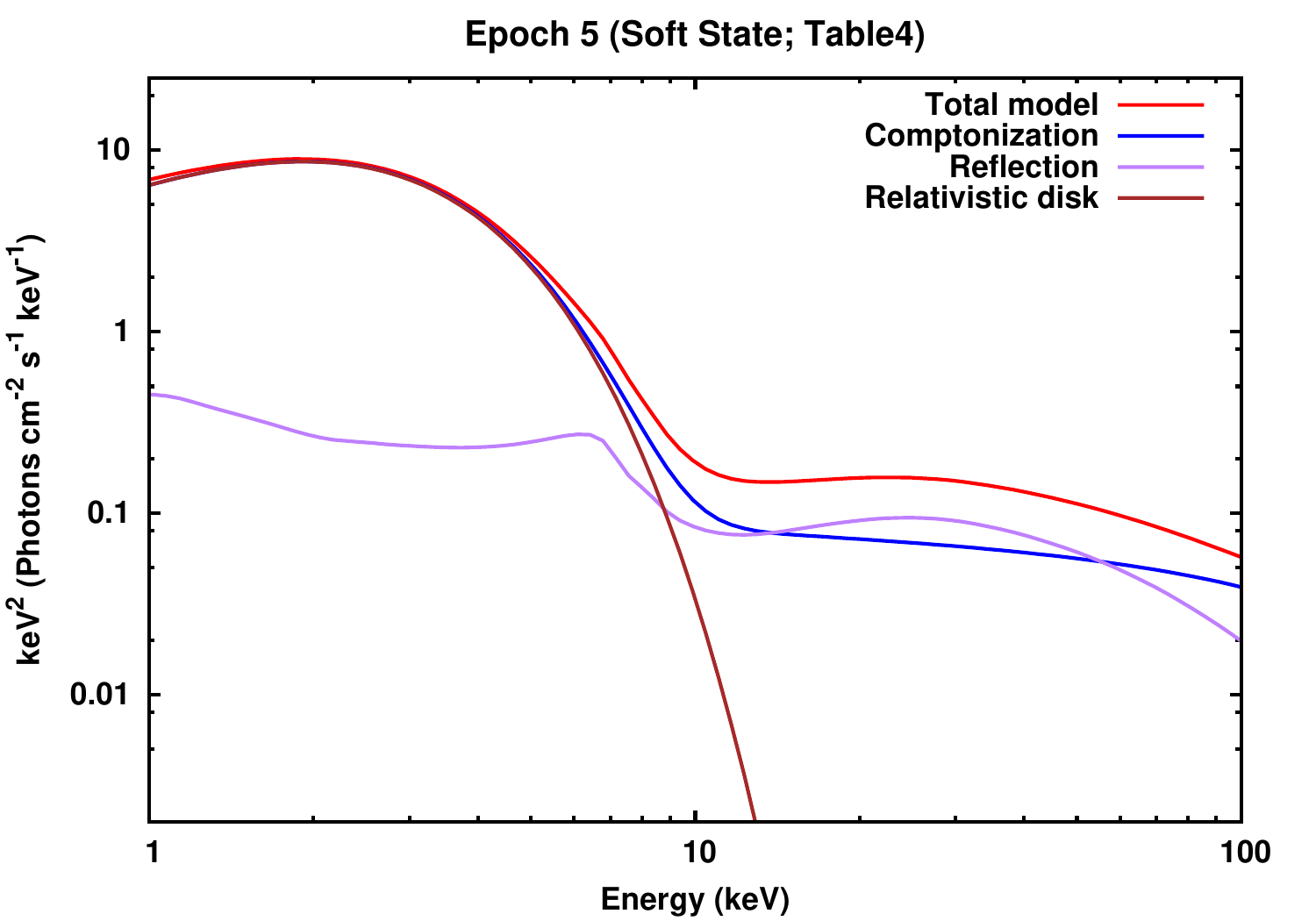}
     }
    \caption{Unabsorbed model components for GX 339–4. The left panel shows the hard state (Epoch 1), and the right panel shows the soft state (Epoch 5). The red curve represents the total model, green indicates warm Comptonization, purple is the reflection component, and brown shows the relativistic disk emission. In the hard state, the blue curve includes both hot and warm Comptonization, while in the soft state, it represents the hot corona alone.}
\label{fig:fig5}
\end{figure*}

In addition to the thermal blackbody emission from the radiatively efficient accretion disk and the nonthermal power-law component resulting from Compton upscattering of disk photons by a hot corona \citep[][]{2004PThPS.155...99Z}, the spectra also exhibit clear signs of reflection. This reflection is a signature of the hard X-ray photons being reprocessed by the accretion disk. Specifically, we used {\fontfamily{pcr}\selectfont relxillCp}, a version of the model that self-consistently incorporates relativistic reflection from a Comptonized component \citep[][]{2014MNRAS.444L.100D, 2014ApJ...782...76G}. The outer radius of the disk was set to 400 $R_g$. The {\fontfamily{pcr}\selectfont relxillCp} model also incorporates a reflection fraction parameter, which represents the ratio of the intensity emitted towards the disk to that escaping to infinity  \citep[][]{2016A&A...590A..76D}. In this analysis, we set the reflection fraction parameter to -1, ensuring that the model only calculates the reflected emission without including additional continuum components from the same source. The accretion disk density was fixed at $10^{18}$ cm$^{-3}$ for all spectral fits. Although the model permits a broken power-law emissivity profile for the reflection component, we adopted a single power-law by tying the inner and outer emissivity indices and fixing their value at 3. For consistency, we tied the photon index ($\Gamma$) and electron temperature ($kT_{\rm e}$) between the {\fontfamily{pcr}\selectfont thcomp} and {\fontfamily{pcr}\selectfont relxillCp} components. The disk inclination and iron abundance ($A_{\rm Fe}$) were kept free. Table~\ref{tab:table2} lists the best-fit spectral parameters for Epoch 1 (hard state) and Epoch 5 (soft state), along with reduced $\chi^2$ values. The black hole spin parameter was not constrained and hence was fixed at the maximum value of 0.998. However, we have verified that fixing the spin at lower values ($\sim 0.5$) does not change the primary results, which are the disc normalization values.

Initially, we applied a single Comptonization model across all epochs. In this setup, we observed a striking discrepancy in disk normalization between Epoch 1 and Epoch 5. Specifically, the disk normalization in the hard state (Epoch 1) was significantly lower than in the soft state (Epoch 5), implying a much smaller inner disk radius in the hard state. This is at odds with the widely accepted view that the accretion disk is truncated at larger radii during the hard state \citep{1997ApJ...489..865E, DoneGierlinskiKubota2007, 2010LNP...794...17G, 2008MNRAS.388..753G, 2015A&A...573A.120P, 2014MNRAS.439.2307F}. To resolve this discrepancy and achieve a more physically consistent interpretation, we introduced a second Comptonization component into the model, representing a warm, optically thick corona. In this updated geometry, the disk photons are first Comptonized by a warm corona, and the resulting photons are subsequently upscattered by a hotter corona. With the inclusion of the warm Comptonization component, the inferred disk normalization in the hard state (Epoch 1) increased significantly compared to the single Comptonization model and became comparable to, or potentially higher than, that in the soft state (Epoch 5). This result aligns more closely with theoretical expectations, where the accretion disk is truncated at larger radii during the hard state. \\
We additionally estimated parameter uncertainties using a Markov Chain Monte Carlo (MCMC) analysis performed with the XSPEC implementation of the Goodman–Weare ensemble sampler.  We used 60 walkers with 50{,}000 steps. The first 3{,}000 steps of each chain were discarded as burn-in. Best-fit values are taken from the peaks of the marginalized posterior distributions. The resulting marginalized constraints are consistent with those obtained from our $\chi^2$-based spectral fitting, with uncertainties quoted at the 90\% confidence level. The MCMC posterior corner plots are presented in Appendix, and the corresponding MCMC-based best-fit values and 90\% confidence uncertainties are listed in Appendix Table~ \ref{tab:table5}. 
In Epoch 5, the spectrum is already well described by only hot Comptonization, and adding a second Comptonization does not lead to any improvement in the fit.
Since \texttt{diskbb} is a phenomenological multi-colour disc model, we performed an additional cross-check using a relativistic thin-disc model to verify that our key results do not depend on the choice of disc model. Specifically, we replaced the \texttt{diskbb} component with the Kerr relativistic disc model \texttt{kerrd} \citep{2003ApJ...597..780E}. The full model used for this test is: \texttt{constant*TBabs(thcomp*kerrd + relxillCp)}. Following standard practice, we fixed the spectral hardening factor to $f_{\rm col}=1.7$ \citep{1995ApJ...445..780S} and set the outer disc radius to $R_{\rm out}=400\,R_g$, where $R_g = GM/c^2$ (e.g., \citet{2025ApJ...984..118D} ). We fixed the black-hole spin to $a_\ast=0.998$, consistent with the high-spin estimates reported for GX~339--4 \citep[e.g.][]{2004ApJ...606L.131M,2008MNRAS.387.1489R,2015ApJ...813...84G,2016ApJ...821L...6P}. The distance to GX~339–4 is uncertain, with lower limits of $\gtrsim 6$ kpc and $\gtrsim 5$ kpc from optical and near-infrared studies \citep{Hynes2003, Heida2017}, and a broader allowed range of $\sim$8–12 kpc from evolutionary modeling \citep{Zdziarski2019}.  In \texttt{relxillCp}, we set the reflection fraction to $-1$ so that the model computes only the reflected emission, and we fixed the disc density to $10^{18}\,\mathrm{cm^{-3}}$ throughout. The inclination and iron abundance were kept free, as in our baseline analysis. To ensure internal consistency, we tied the photon index and electron temperature between \texttt{thcomp} and \texttt{relxillCp}. Additionally, the inclination and inner disc radius in \texttt{relxillCp} were tied to those of \texttt{kerrd}. We fix the distance at $D = 6$ kpc. We first fitted the soft-state spectrum (Epoch~5) with the inner disc radius fixed at the ISCO and obtained an acceptable fit ($\chi^2_\nu=1.275$), yielding a black-hole mass of $M_{\rm BH}=17.54^{+0.90}_{-0.73}\,M_\odot$ (90\% confidence). We then fixed the mass to $M_{\rm BH}=17.5\, M_\odot$ for the hard-state fit (Epoch~1). Note the the black hole mass depends on the assumed distance of $D = 6$ kpc. Best-fit parameters from the \texttt{kerrd}-based fits are reported in Table~\ref{tab:table6}, along with $\chi^{2}/\mathrm{dof}$ for Epochs~1 and~5. In the hard state, the single-Comptonization model gives $R_{\rm in}<1.62$ with $\chi^2_\nu=1.46$. However, adding a warm Comptonization component (dual-Comptonization) significantly improves the fit ($\chi^2_\nu=0.94$) and yields a larger inner radius, $R_{\rm in}=14.1^{+1.8}_{-2.2}$ (90\% confidence), consistent with a truncated disc in the hard state. These results confirm that the main inference of our work is robust to replacing \texttt{diskbb} with a relativistic thin-disc prescription.\\
To ensure consistency across all epochs, we examined model parameters that are expected to remain constant throughout the outburst. The iron abundance, being an intrinsic characteristic of the disk, is not expected to vary between observations. Likewise, the inclination angle should remain unchanged throughout the outburst. Therefore, we first allowed these parameters to vary freely during the spectral fits across all epochs to determine their best-fit values. We then calculated the weighted average of these values and fixed the averages in all final spectral fits.

We began by applying the single Comptonization model to all five epochs. From spectral fits, we obtained an average inclination angle of $30^\circ$ and an average iron abundance of $A_{\rm Fe} = 9$. These values were then fixed in the single Comptonization model used for the final fits. The corresponding best-fit spectral parameters are summarized in Table~\ref{tab:table3}. In the hard state (Epochs 1 and 2), the disk normalization is notably lower, with values of 0.38 and 0.43 (in units of $10^3$), respectively. In contrast, during the soft state (Epochs 4 and 5), the normalization increases to 2.95 and 2.92. This indicates a much smaller apparent inner disk radius in the hard state. Figure \ref{fig:fig3} presents the simultaneous broadband X-ray spectra of GX 339-4 during the hard (Epoch 1) and soft (Epoch 5) states, fitted with the single-Comptonization model (Table~3).

\begin{figure}
    \centering
	\includegraphics[width=1.\columnwidth]{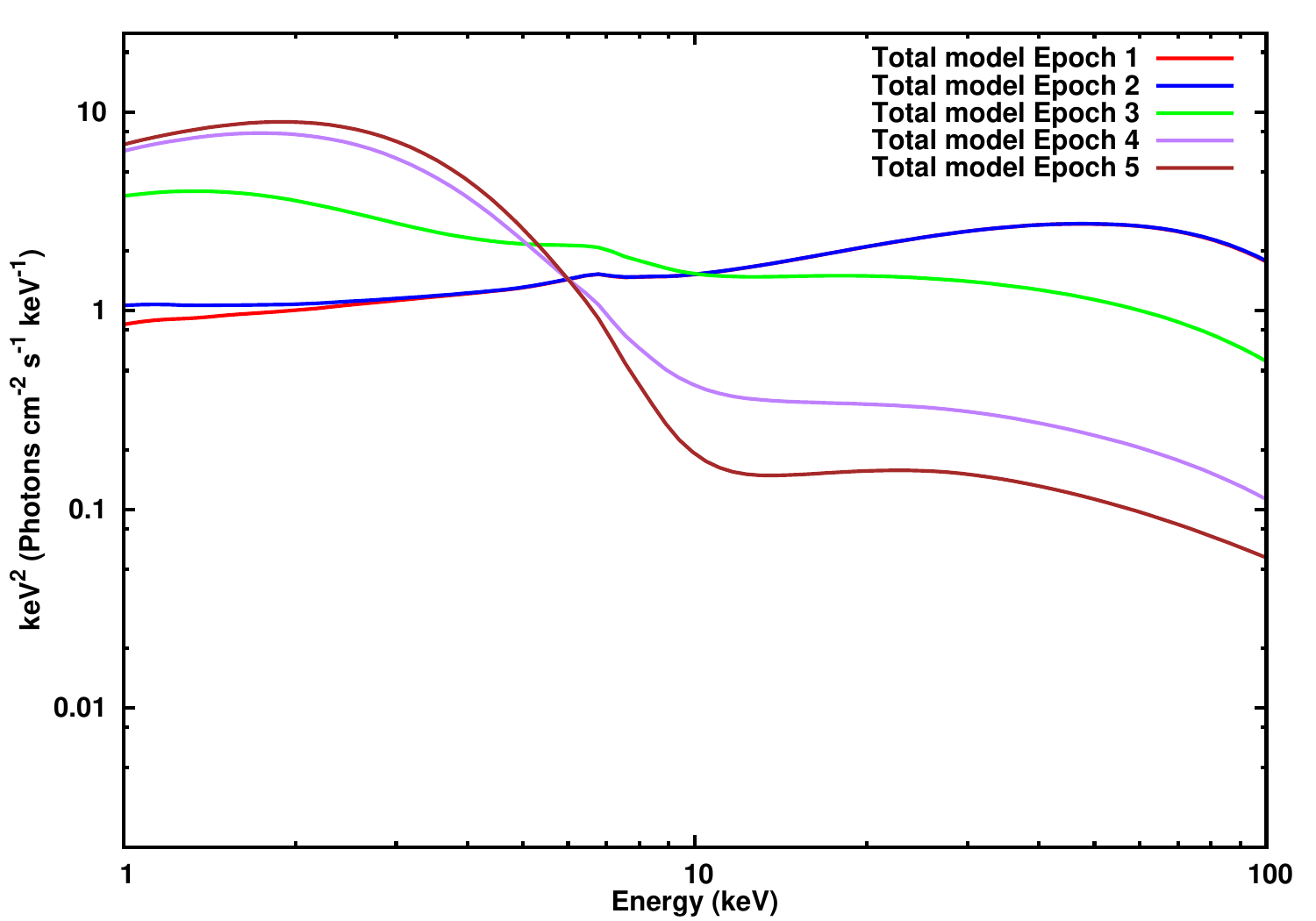}
        \caption{The figure displays the unabsorbed total model using dual Comptonization for Epochs 1, 2, 3, and 4 and single Comptonization for epoch 5, illustrating the transition of the source from the hard state to the soft state.}
    \label{fig:fig6}
\end{figure}

We then applied the dual Comptonization model to all five epochs to further investigate the evolution of the disk normalization. Following the same approach as in the single Comptonization fits, we first allowed the inclination angle and iron abundance to vary across all epochs to determine their average values. The average inclination angle was found to be $22.6^\circ$, and the average iron abundance was $A_{\rm Fe} = 8.3$. These values were then fixed in the final spectral fits for consistency. The resulting best-fit parameters for all epochs using the dual Comptonization model are presented in Table \ref{tab:table4}. Figure \ref{fig:fig4} presents the simultaneous broadband X-ray spectra of GX 339-4 during the hard (Epoch 1) and soft (Epoch 5) states, fitted with the best-fitting model(s) reported in Table \ref{tab:table4} and \ref{fig:fig5} show the unabsorbed model components for these epochs. Figure \ref{fig:fig6} shows the evolution of the unabsorbed total model spectra across Epochs 1–5.  \\
In Epoch 5 (softest state) the spectrum is already well described without a warm component. When the iron abundance ($A_{\rm Fe}$) and inclination angle ($i$) are allowed to be free (Table \ref{tab:table2}), the single-Comptonization fit gives $\chi^2/{\rm dof}=677.4/642$, and adding a second Comptonization does not lead to any improvement in the fit. However, the best-fit inclination angles are significantly different for the hard state (Epoch 1) and the soft state (Epoch 5) for a single Comptonization fit (Table \ref{tab:table2}). Thus, when we fix the inclination angle and the Iron abundance to be consistent (i.e. $i=22.6^\circ$, $A_{\rm Fe} = 8.3\,A_{\rm Fe,\odot}$), the fit for Epoch 5 considerably worsens and gives $\chi^2/{\rm dof} = 728.9/644$. On the other hand, the best-fit inclination angle for the dual Comptonization model in the soft state (Epoch 5) is consistent with that obtained for Epoch 1. Hence, when inclination angle and Iron abundance are fixed at $i=22.6^\circ$ and  $A_{\rm Fe} = 8.3\,A_{\rm Fe,\odot}$, the fitting using a single Comptonization model for Epoch 1 does not become worse. We note that while the soft state spectrum of Epoch 5 does not seem to require a warm Comptonization component, the fitting for the other soft state spectrum (Epoch 4) does improve with the addition.

\begin{figure}
    \centering
	\includegraphics[width=1.\columnwidth]{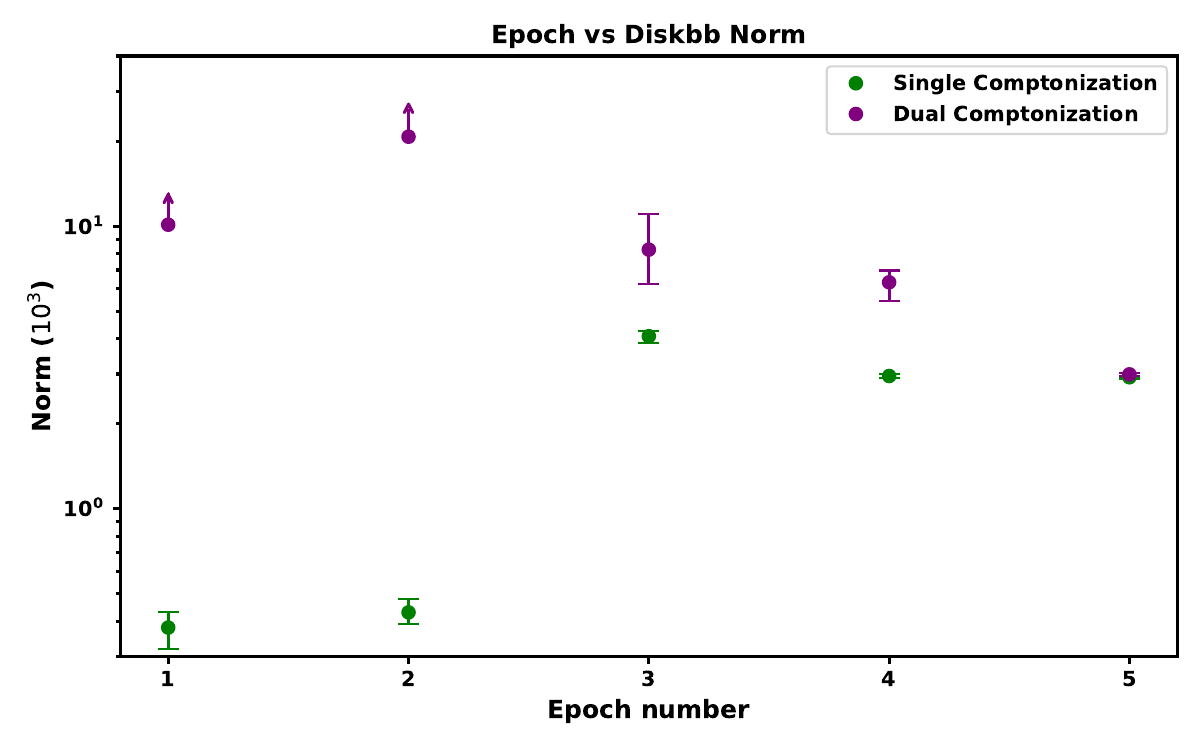}
\caption{Evolution of \textit{diskbb} normalization parameters across observational epochs for both single and dual Comptonization models. Green points with error bars represent the single Comptonization model, while purple points represent the dual Comptonization scenario. Upward arrows in Epochs 1 and 2 of the dual Comptonization model indicate lower limits, where only a lower bound is constrained. Error bars correspond to 90\% confidence level uncertainties.}
    \label{fig:fig7}
\end{figure}

The evolution of the disk normalization across the epochs, as compared between the single and dual Comptonization models, is shown in Figure~\ref{fig:fig7}. In the single Comptonization case, the disk normalization is significantly lower during the hard state compared to the soft state. This leads to a smaller inferred inner disk radius in the hard state, which contradicts the standard picture of a truncated accretion disk that is expected to lie at larger radii during this state. In contrast, the dual Comptonization model yields significantly higher disk normalization values in Epochs 1 and 2, exceeding more than $20 \times 10^3$ and $50 \times 10^3$, respectively. These higher values are more consistent with a physically truncated, cooler, and less luminous disk in the hard state. This comparison highlights how the inclusion of a warm corona alters the inferred disk properties and leads to a more physically realistic interpretation of the accretion geometry.

To further assess whether the soft disc component is statistically required in the hard-state spectrum, we performed an additional test without an explicit disc component. Since \texttt{thcomp} is used as a convolution applied to the disc seed spectrum in our baseline model, we replaced \texttt{thcomp*diskbb} with an additive Comptonization component \texttt{nthcomp} which essentially implies that the covering fraction is unity and that we are only observing the Comptonized component. The resulting hard-state model \texttt{constant*TBabs(nthcomp + relxillCp)} provides a significantly worse description of the data, with the fit degrading from $\chi^2_\nu \simeq 1.04$ (with \texttt{thcomp*diskbb)} to $\chi^2_\nu \simeq 1.47$ (with \texttt{nthcomp}), and clear low-energy residuals remaining. This indicates that a distinct disk component is still statistically required in the hard-state spectrum for our dataset.

\section{Discussion}
\label{sec:discussion}

The geometry of the accretion flow in BHXBs, particularly the location of the inner edge of the accretion disk across different spectral states, remains a subject of ongoing debate. A key question is whether the disk is truncated at larger radii in the hard state or extends down to the innermost stable circular orbit (ISCO), as is typically observed in the soft state. In this work, we performed a detailed spectral analysis of GX 339–4 using simultaneous NICER, NuSTAR, and HXMT observations that span both hard and soft states. Our primary goal was to better understand the behavior of the inner accretion disk during the hard state and across the spectral transition.

Previous studies have reported that the inner edge of the accretion disk moves inward as the spectrum softens. For instance, \citet{2010MNRAS.407.2287D} and \citet{2015ebha.confE..71B} observed that the relativistic Fe K$\alpha$ line becomes increasingly broadened during the transition from hard to soft states. This is attributed to the disk extending closer to the black hole, amplifying relativistic effects near the ISCO and resulting in broader and more skewed line profiles.

Using the single Comptonization model (Table~\ref{tab:table3}), we found that the disk normalization is lower in the hard state than in the soft state. This result implies a smaller inner disk radius in the hard state, which contradicts the standard scenario where the disk is expected to be truncated at larger radii during this state and move inward as the source transitions to the soft state.

A similar inconsistency was reported recently by \citet{Li2025}, who analyzed 2018 NICER and \textit{Insight}-HXMT observations of MAXI J1820+070. They applied a single Comptonization model and found that the inner disk radius reached a minimum of approximately 4.5 $R_g$ during the intermediate state, which was smaller than the radius in the soft state (around 9 $R_g$). Their result supports our observation that the single Comptonization model tends to underestimate the inner radius in states outside the soft state. \\
As an independent consistency check, we compare our disk radius 
estimate with that of \citet{2024MNRAS.527.2128J}. One observation is common to both studies, corresponding to MJD~59308. Using the \texttt{diskbb} normalization reported in Table~3 and adopting the same spectral hardening factor as used by \citet{2024MNRAS.527.2128J}, we estimate the inner disc radius for this observation. Assuming $\kappa = 1.8$, an inclination angle of $\theta = 30^\circ$, and a source distance of $d = 8.4 \pm 0.9$ kpc \citep{Parker2016}, we obtain $R_{\mathrm{in}} \approx 75$ km. Jana et al.\ (2024) reported an inner disc radius of $\sim 45$ km for the same observation. This discrepancy may arise from the use of different spectral models, instruments, and associated calibration uncertainties.

To resolve the unphysical result of a smaller disk radius in the hard state than in the soft state, we incorporated a second Comptonization component into our model. This additional component represents a warm, optically thick corona located close to the disk that upscatters thermal photons to produce soft Comptonized emission. In the single Comptonization model, the hard-state disk normalization remained smaller than in the soft state. However, with the inclusion of the warm Comptonization component, the disk normalization in the hard state increased significantly, bringing it more in line with the expected geometry of a truncated disk. The dual-corona geometry, which consists of a warm corona and a hot, optically thin corona responsible for the hard X-ray emission, is physically motivated and supported by several previous studies \citep[][]{yamada2013, basak2017, zdziarski2021, chand2024}. These studies have demonstrated that this configuration provides a more accurate description of spectral features in the hard state, where the accretion disk may still be truncated. The warm Comptonization component invoked here is physically analogous to the warm corona used to explain the soft X-ray excess in AGN \citep[][]{Done2012, Petrucci2018}. However, because the accretion disc in GX 339$-$4 already peaks in the soft X-ray band, the warm corona does not produce a distinct spectral excess as in AGN, but instead manifests as a modification of the apparent disc normalization.

While the primary focus of this work was to investigate the evolution of disk normalization, we also observed systematic changes in other spectral parameters across the hard-to-soft transition. 
In the soft state, the contribution of the Comptonized component is weak, and the spectral index is steep, resulting in the electron temperature of the hot corona being poorly constrained and physically inconclusive, as also evident from Figure \ref{fig:fig9}. Similar difficulties in constraining coronal cutoffs in soft states, due to weak high-energy tails, have been reported previously \citep{ZdziarskiGierlinski2004,DoneGierlinskiKubota2007}. The photon index ($\Gamma$) softens from 1.62 to 2.14 during the hard-to-soft transition. Concurrently, the scattering fraction decreases sharply from approximately 0.54 to 0.01, indicating a diminishing influence of the Comptonizing corona and a growing dominance of the thermal disk emission as the source softens. The disk also becomes hotter across the transition, with the inner disk temperature ($T_{\rm in}$) rising from around 0.27 keV in Epoch 1 to 0.81 keV in Epoch 5. Importantly, the disk normalization is higher in the hard state (with the dual Comptonization model), consistent with a cooler, truncated disk that appears more extended in the early stages of the outburst. Our results are consistent with the standard understanding of accretion evolution, in which the inner disk moves closer to the ISCO as the spectrum softens \citep{1997ApJ...489..865E,RemillardMcClintock2006,DoneGierlinskiKubota2007}. However, this expected trend becomes apparent only when a dual-corona model is used, highlighting the importance of including both hot and warm Comptonization components in spectral modeling. The inner disk radius inferred from the reflection component ($R_{\rm in}$) decreases from approximately 10.4 $R_g$ in Epoch 1 to less than 2.7 $R_g$ in Epoch 5, indicating that the disk reaches the ISCO in the soft state. These results support a dual-corona geometry in GX 339–4, where including a warm Comptonization component yields physically consistent disk parameters in the hard state. This approach helps avoid underestimating the inner disk radius and better captures the accretion geometry across spectral states. Future studies should explore these parameter variations in a broader sample of black hole X-ray binaries, using high-sensitivity and multi-wavelength observations to improve our understanding of spectral state transitions.

%%%%%%%%%%%%%%%%%%%%%%%%%%%%%%%%%%%%%%%%%%%%%%%%%%%%%%%%%%%%%%%%%%%%%%%%%%%%%%%%%%%%%

\section*{Acknowledgements}
\label{sec:acknowledgement}
We would like to thank the anonymous reviewer for constructive comments. We acknowledge the utilization of software and resources provided by the High Energy Astrophysics Science Archive Research Center (HEASARC), a service of NASA's Astrophysics Science Division at NASA/GSFC, and the High Energy Astrophysics Division of the Smithsonian Astrophysical Observatory.
We acknowledge the use of the NuSTAR Data Analysis Software (NUSTARDAS), a collaborative development between the ASI Space Science Data Center (SSDC, Italy) and the California Institute of Technology (Caltech, USA).
The author, RD, acknowledges the valuable visits to the Inter-University Centre for Astronomy and Astrophysics (IUCAA), Pune, where a significant portion of the work was completed.

\section*{Data Availability}
\label{sec:data_availability}

The data used in this article are publicly available from their respective mission archives. NICER and NuSTAR data can be accessed via the HEASARC Browse interface (\url{https://heasarc.gsfc.nasa.gov/db-perl/W3Browse/w3browse.pl}). HXMT data are available from the Insight-HXMT Data Archive at \href{http://hxmten.ihep.ac.cn/}{http://hxmten.ihep.ac.cn/}. MAXI data can be obtained from the MAXI website (\url{http://maxi.riken.jp/top/index.html}).

%%%%%%%%%%%%%%%%%%%%%%%%%%%%%%%%%%%% REFERENCES %%%%%%%%%%%%%%%%%%%5%%%%%%%%%%%%%%%

% The best way to enter references is to use BibTeX:

\bibliographystyle{mnras}
\bibliography{example_new}

%%%%%%%%%%%%%%%%%%%%%%%%%%%%%%%%%%%%%%%%%%%%%%%%%%

%%%%%%%%%%%%%%%%% APPENDICES %%%%%%%%%%%%%%%%%%%%%

%%%%%%%%%%%%%%%%%%%%%%%%%%%%%%%%%%%%%%%%%%%%%%%%%%

%\section{Appendix}
\appendix
% Requires: \usepackage{booktabs} \usepackage{multirow}

\section{APPENDIX}

\begin{figure*}
\vspace{-10pt} % reduce space above the figure
     \subfloat{
         \includegraphics[width=0.49\textwidth]{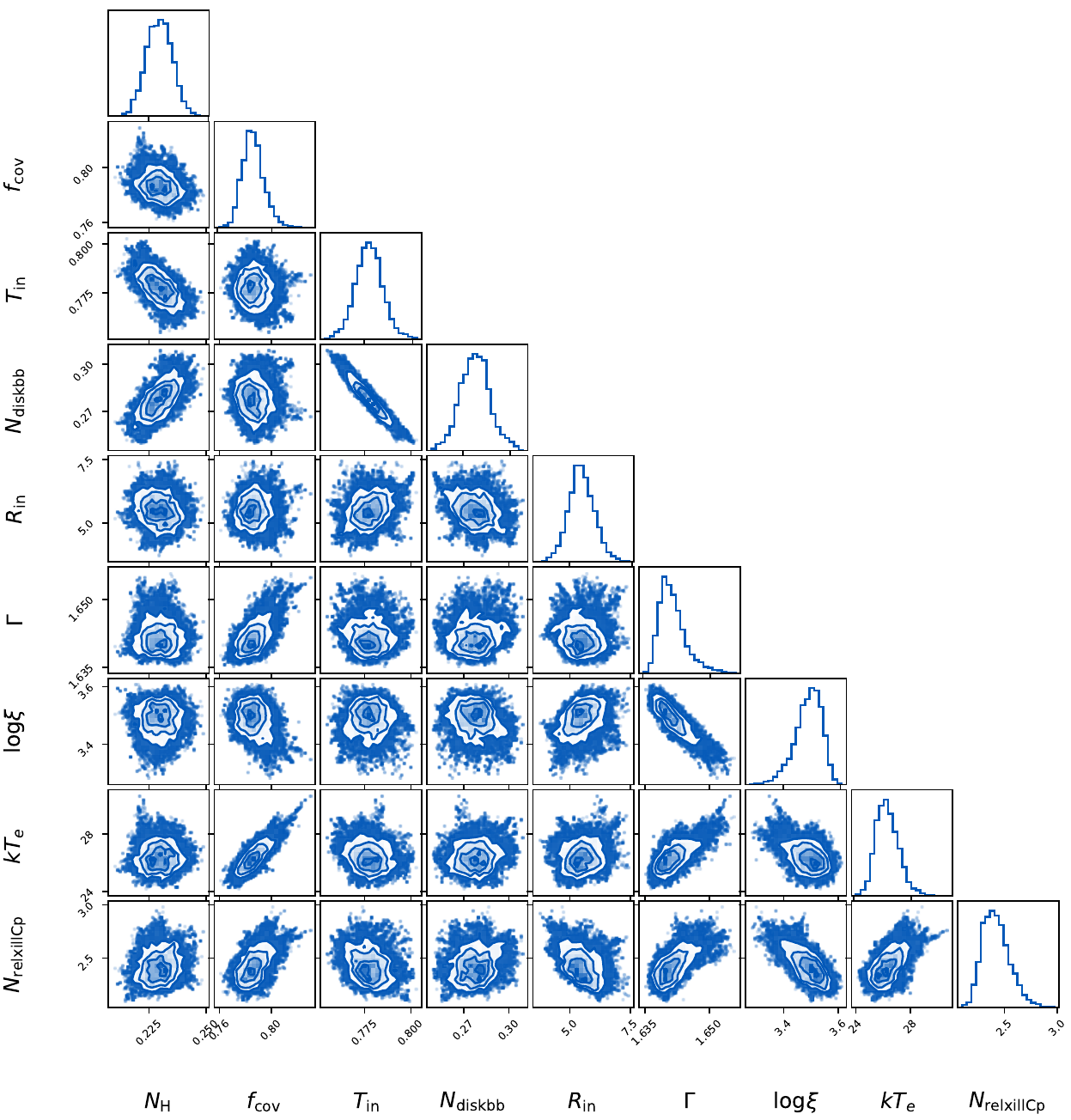}
     }
     \hfill
     \subfloat{
         \includegraphics[width=0.49\textwidth]{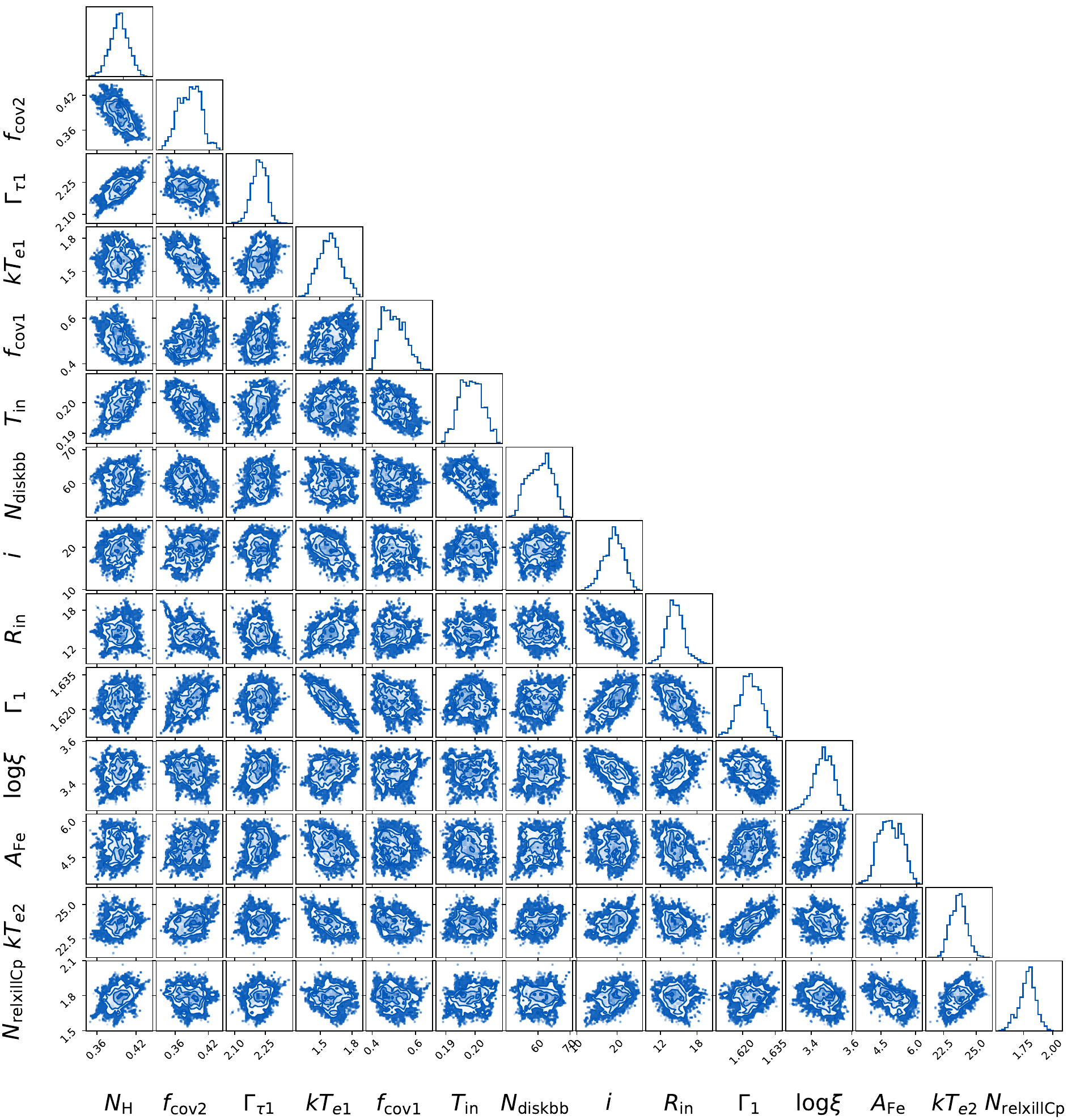}
     }
     
\caption{Corner plots of spectral parameters obtained from the MCMC analysis for Epoch~1:
The hot Comptonization model (left) and the hot+warm Comptonization model (right).
The one-dimensional histograms show the marginalized posterior probability
distributions, normalized such that the area under each curve is unity.
The normalization of \texttt{relxillCp}, $N_{\rm relxillCp}$, is expressed in units
of $10^{-3}$, while the disk blackbody normalization, $N_{\rm diskbb}$, is expressed
in units of $10^{3}$. All other parameters are quoted in the units listed in Table~\ref{tab:table5}. For the hot+warm Comptonization model (right), $kT_{e,1}$, $f_{{\rm cov},1}$, and $\Gamma_1$
correspond to the warm corona, while $kT_{e,2}$, $f_{{\rm cov},2}$, and $\Gamma_2$
correspond to the hot corona.
}

\label{fig:fig8}
\end{figure*}

\begin{figure}
    \centering
	\includegraphics[width=1.\columnwidth]{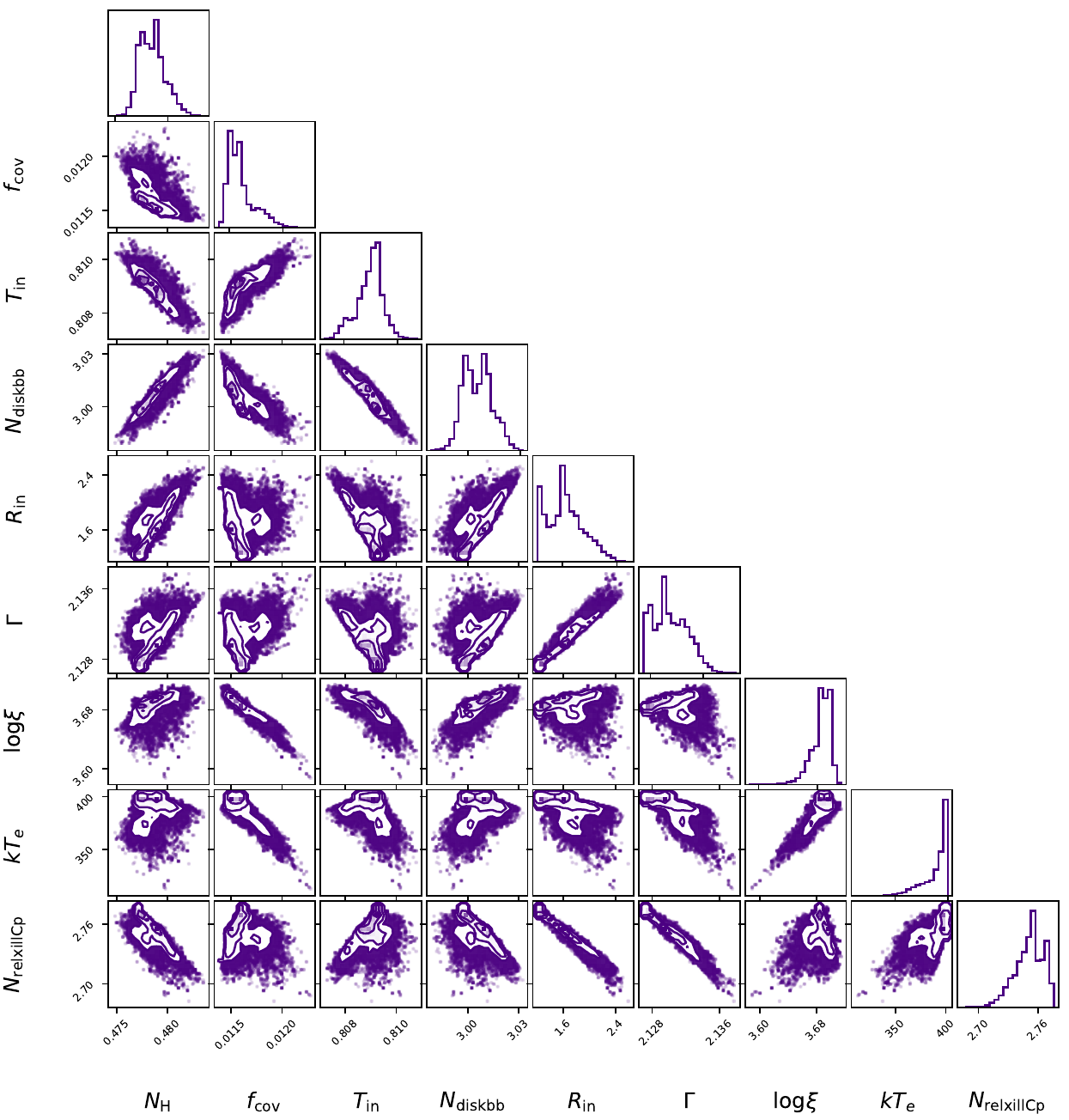}
\caption{Corner plots of spectral parameters obtained from the MCMC analysis for Epoch~5
using the single Comptonization model.
The one–dimensional histograms show the marginalized posterior probability
distributions, normalized such that the area under each curve is unity.
The normalization of \texttt{relxillCp}, $N_{\rm relxillCp}$, is expressed in units
of $10^{-3}$, while the disk blackbody normalization, $N_{\rm diskbb}$, is expressed
in units of $10^{3}$. All other parameters are quoted in the units listed in Table~\ref{tab:table5}. }
    \label{fig:fig9}
\end{figure}

\begin{table*}
\setlength{\tabcolsep}{10pt}
\centering
\caption{Best-fit spectral parameters. Error bars are derived from MCMC posterior distributions and quoted as 90\% confidence intervals.}
\label{tab:table5}
%\begin{tabular}{ M{2.7cm} M{1.7cm} M{2.4cm} M{2.4cm} M{2.3cm} M{2.cm} }

\begin{adjustbox}{max width=\textwidth}
\begin{tabular}{c c c c c}
\hline\hline
Components & Parameters  
& Epoch 1 (Single thcomp) 
& Epoch 1 (Dual thcomp) 
& Epoch 5 (Single thcomp) \\[2pt]
\hline

Tbabs 
& $N_H$ ($10^{22}\,\mathrm{cm}^{-2}$)
& $0.23^{+0.01}_{-0.01}$
& $0.40^{+0.02}_{-0.02}$
& $0.480^{+0.001}_{-0.003}$ \\[2pt]
\hline

Hard Comptonization 
& $\Gamma$
& $1.64^{+0.01}_{-0.01}$
& $1.62^{+0.01}_{-0.01}$
& $2.13^{+0.01}_{-0.01}$ \\[2pt]

& $kT_e$ (keV)
& $25.7^{+2.1}_{-0.4}$
& $23.5^{+1.1}_{-0.7}$
& $>300$ \\[2pt]

& $f_{\rm SC}$
& $0.78^{+0.01}_{-0.01}$
& $0.39^{+0.03}_{-0.03}$
& $0.0114^{+0.0001}_{-0.0002}$ \\[2pt]
\hline

Soft Comptonization 
& $\Gamma$
& --
& $2.22^{+0.06}_{-0.05}$
& -- \\[2pt]

& $kT_{\rm e}$ (keV)
& --
& $1.6^{+0.2}_{-0.2}$
& -- \\[2pt]

& $f_{\rm SC}$
& --
& $0.49^{+0.10}_{-0.06}$
& -- \\[2pt]
\hline

Disk 
& $T_{\rm in}$ (keV)
& $0.78^{+0.02}_{-0.01}$
& $0.20^{+0.01}_{-0.01}$
& $0.809^{+0.001}_{-0.000}$ \\[2pt]

& Norm. ($10^{3}$)
& $0.28^{+0.02}_{-0.02}$
& $60^{+5}_{-6}$
& $3.02^{+0.01}_{-0.01}$ \\[2pt]
\hline

Reflection 
& Incl. (deg)
& $28.6^{+2.2}_{-2.2}$
& $19.7^{+3.0}_{-4.6}$
& $19.5^{+2.1}_{-0.0}$ \\[2pt]

& $R_{\rm in}$ ($r_g$)
& $5.3^{+1.1}_{-0.6}$
& $14.3^{+2.6}_{-1.9}$
& $1.5^{+0.7}_{-0.2}$ \\[2pt]

& $\log\xi$ ($\mathrm{erg\,cm\,s^{-1}}$)
& $3.52^{+0.03}_{-0.12}$
& $3.46^{+0.07}_{-0.08}$
& $3.70^{+0.01}_{-0.03}$ \\[2pt]

& $A_{\rm Fe}$
& $10.0^{+1.1}_{-0.6}$
& $5.0^{+0.7}_{-0.8}$
& $>7.4$ \\[2pt]

& Norm. ($10^{-3}$)
& $2.29^{+0.36}_{-0.05}$
& $1.74^{+0.15}_{-0.09}$
& $2.78^{+0.03}_{-0.93}$ \\[2pt]
\hline
& $\chi^2/\mathrm{bins}$ (FPMA) & 290.7/264 & 278.4/264 & 254.0/214 \\[2pt]
& $\chi^2/\mathrm{bins}$ (FPMB) & 275.3/268 & 253.9/268 & 266.6/218 \\[2pt]
& $\chi^2/\mathrm{bins}$ (LE)   & 62.3/50   & 50.6/50   & 61.3/54  \\[2pt]
& $\chi^2/\mathrm{bins}$ (ME)   & 12.3/14   & 13.4/14   & 7.1/14   \\[2pt]
& $\chi^2/\mathrm{bins}$ (HE)   & 4.4/14    & 6.4/14    & 9.0/15   \\[2pt]
& $\chi^2/\mathrm{bins}$ (XTI)  & 121.5/142 & 86.7/142  & 79.3/143 \\[2pt]
& $\chi^2/\mathrm{dof}$ (Total) & 766.5/736 & 689.3/733 & 677.4/642 \\[2pt]

\hline\hline
\end{tabular}
\end{adjustbox}
\end{table*}

%%%%%%%%%%%%%%%%%%%%%%%%%%%%%%%%%%%%%%%%%%%%%%%%%%%%%%%%%%%%%%%%%%%%%%%%%%%%%
\begin{table*}
\setlength{\tabcolsep}{10pt}
\centering
\caption{Broadband X-ray spectral parameters of GX~339--4 from simultaneous NICER, NuSTAR, and Insight-HXMT observations, analyzed using single and dual Comptonization models with a relativistic disc component (\texttt{kerrd}) and reflection.}
\label{tab:table6}
\begin{tabular}{c c c c c}
\hline \hline
Components & Parameters &
Epoch 1 (Single thcomp) &
Epoch 1 (Dual thcomp) &
Epoch 5 (Single thcomp) \\ [2pt]
\hline
Tbabs
& $N_H$ ($10^{22}\,\mathrm{cm^{-2}}$)
& $0.376^{+0.007}_{-0.006}$
& $0.399^{+0.055}_{-0.071}$
& $0.535^{+0.006}_{-0.005}$ \\ [2pt]

Hard Comptonization
& $\Gamma$
& $1.670^{+0.004}_{-0.001}$
& $1.623^{+0.006}_{-0.015}$
& $2.476^{+0.068}_{-0.029}$ \\ [2pt]

& $kT_e$ (keV)
& $50.0^{+0.6}_{-0.8}$
& $23.48^{+1.06}_{-1.05}$
& $>149$ \\ [2pt]

& $f_{\mathrm{SC}}$
& $0.871^{+0.007}_{-0.028}$
& $0.345^{+0.020}_{-0.118}$
& $0.0261^{+0.0008}_{-0.0029}$ \\ [2pt]

Soft Comptonization
& $\Gamma$
& --
& $2.215^{+0.057}_{-0.072}$
& -- \\ [2pt]

& $kT_e$ (keV)
& --
& $1.60^{+0.30}_{-0.18}$
& -- \\ [2pt]

& $f_{\mathrm{SC}}$
& --
& $0.428^{+0.233}_{-0.168}$
& -- \\ [2pt]

Relativistic disk
& $\dot{M}$ ($10^{18} \, \text{gm} \, \text{s}^{-1}$)
& $0.0283^{+0.0011}_{-0.0006}$
& $0.072^{+0.055}_{-0.041}$
& $0.591^{+0.002}_{-0.002}$ \\  [2pt]

Reflection
& Incl. (deg)
& $39.0^{+1.2}_{-0.8}$
& $19.7^{+4.0}_{-2.5}$
& $44.16^{+0.81}_{-0.43}$ \\ [2pt]

& $R_{\mathrm{in}}$ ($r_g$)
& $<1.62$
& $14.1^{+1.8}_{-2.2}$
& $1.235^{\mathrm{f}}$  \\ [2pt]

& $\log \xi$ ($\mathrm{erg\,cm\,s^{-1}}$)
& $3.36^{+0.02}_{-0.05}$
& $3.46^{+0.08}_{-0.19}$
& $>0.82$ \\ [2pt]

& $A_{\mathrm{Fe}}$
& $4.93^{+0.15}_{-0.32}$
& $5.03^{+1.84}_{-2.05}$
& $>9.66$ \\ [2pt]

& Norm. ($10^{-3}$)
& $5.26^{+0.19}_{-0.19}$
& $1.75^{+0.40}_{-0.15}$
& $14.72^{+1.79}_{-1.68}$ \\ [2pt]

\hline
& $\chi^2/\mathrm{bins}$ (FPMA)
& 487.83/264 & 278.30/264 & 309.9/214 \\ [2pt]
& $\chi^2/\mathrm{bins}$ (FPMB)
& 416.99/268 & 253.39/268 & 294.3/218 \\ [2pt]
& $\chi^2/\mathrm{bins}$ (LE)
& 45.26/50 & 51.18/50 & 56.9/54 \\ [2pt]
& $\chi^2/\mathrm{bins}$ (ME)
& 12.01/14 & 13.34/14 & 7.2/14 \\ [2pt]
& $\chi^2/\mathrm{bins}$ (HE)
& 26.54/14 & 6.42/14 & 7.8/15 \\ [2pt]
& $\chi^2/\mathrm{bins}$ (XTI)
& 88.14/142 & 86.78/142 & 123.7/143 \\ [2pt]
& $\chi^2/\mathrm{dof}$ (Total)
& 1076.77/737 & 689.42/734 & 799.9/643 \\ [2pt]
\hline \hline

\multicolumn{5}{c}{\footnotesize $^{\mathrm{f}}$ Parameter frozen during the fitting.} \\[-2pt]
\end{tabular}
\end{table*}

%%%%%%%%%%%%%%%%%%%%%%%%%%%%%%%%%%%%%%%%%%%%%%%%%%%%%%%%%%%%%%%%%%%%%%%%%%%%%

\label{lastpage}
\end{document}